\documentstyle[11pt,aaspp4]{article}
\journalid{}{}
\articleid{}{}
\slugcomment{}
\begin{document}

\title{Galactic-Scale Outflow and Supersonic Ram-Pressure Stripping in the 
Virgo Cluster Galaxy NGC~4388}

%\title{NGC~4388: Galactic-Scale Nuclear Outflow and Supersonic Interaction
%with the Virgo Intracluster Medium}
%\title{Origin of the Extraplanar Gas in NGC 4388: Galactic Outflow or
%Ram-Pressure Stripping by the Virgo Cluster Intracluster Medium?}
%\title{Origin of the Extended Ionized Gas in NGC 4388: Galactic Outflow or
%Ram-Pressure Stripping by the Virgo Cluster Intracluster Medium?}
%\title{Origin of the Ionized Gas in the Halo of the Virgo Cluster (Core)
%Galaxy NGC 4388: Galactic Outflow or Ram-Pressure Stripping?}
%\title{Origin of the Ionized Gas in the Halo of the Virgo Cluster (Core)
%Galaxy NGC 4388: Galactic Outflow or Ram-Pressure Stripping?}

\author{Sylvain Veilleux\altaffilmark{1,2}, Jonathan
Bland-Hawthorn\altaffilmark{3}, Gerald Cecil\altaffilmark{4,5}, R. Brent
Tully\altaffilmark{6}, \\
and Scott T. Miller\altaffilmark{1}}

\altaffiltext{1}{Department of Astronomy, University of Maryland,
College Park, MD 20742; E-mail: veilleux@astro.umd.edu}

\altaffiltext{2}{Cottrell Scholar of the Research Corporation}

\altaffiltext{3}{Anglo-Australian Observatory, P.O. Box 296, Epping, 
NSW 2121, Australia; E-mail: jbh@aaoepp2.aao.gov.au}

\altaffiltext{4}{Department of Physics and Astronomy, University of North
Carolina, CB\# 3255, Chapel Hill, NC 27599-3255; E-mail: cecil@noao.edu}

\altaffiltext{5}{SOAR Telescope Project, NOAO, 950 N. Cherry Ave.,
Tucson, AZ 85726-6732}

\altaffiltext{6}{Institute for Astronomy, University of Hawaii, 2680
Woodlawn Drive, Honolulu, HI 96822; E-mail: tully@galileo.ifa.hawaii.edu}

\begin{abstract}
The Hawaii Imaging Fabry-Perot Interferometer (HIFI) on the University
of Hawaii 2.2m telescope was used to map the H$\alpha$ and [O~III]
$\lambda$5007 emission-line profiles across the entire disk of the
edge-on Sb galaxy NGC~4388.  We confirm a rich complex of highly
ionized gas that extends $\sim$ 4 kpc above the disk of this galaxy.
Low-ionization gas associated with star formation is also present in
the disk. Evidence for bar streaming is detected in the disk component
and is discussed in a companion paper (Veilleux, Bland-Hawthorn, \&
Cecil 1999; hereafter VBC).  Non-rotational blueshifted velocities of
50 $-$ 250 km s$^{-1}$ are measured in the extraplanar gas north-east
of the nucleus.  The brighter features in this complex tend to have
more blueshifted velocities. A redshifted cloud is also detected 2 kpc
south-west of the nucleus. The velocity field of the extraplanar gas
of NGC~4388 appears to be unaffected by the inferred supersonic (Mach
number $M$ $\approx$ 3) motion of this galaxy through the ICM of the
Virgo cluster. We argue that this is because the galaxy and the
high-$\vert$z$\vert$ gas lie behind a Mach cone with opening angle
$\sim$ 80$\arcdeg$.  The shocked ICM that flows near the galaxy has a
velocity of $\sim$ 500 km s$^{-1}$ and exerts insufficient ram
pressure on the extraplanar gas to perturb its kinematics.  We
consider several explanations of the velocity field of the extraplanar
gas. Velocities, especially blueshifted velocities on the N side of
the galaxy, are best explained as a bipolar outflow which is tilted by
$>$12$\arcdeg$ from the normal to the disk.  The observed offset
between the extraplanar gas and the radio structure may be due to
buoyancy or refractive bending by density gradients in the halo gas.
Velocity substructure in the outflowing gas also suggests an
interaction with ambient halo gas.

\end{abstract}


\section{Introduction}

The edge-on Sb galaxy NGC 4388 is projected
$<$1$\arcdeg$ from the core of the Virgo cluster (Biggnelli, Tammann, \&
Sandage 1987; Yasuda, Fukugita, \& Okamura 1997), but its systemic
velocity of $\sim$ 2,525 km s$^{-1}$
%Helou et al. 1979, 1981; Ayani \&
%Iye 1989; Corbin, Baldwin, \& Wilson 1988; Petitjean \& Durret 1993;
%Rubin, Kenney, \& Young 1997) 
is $\sim$ 1,500 km s$^{-1}$ above that of the Virgo cluster (e.g.,
Rubin, Kenney, \& Young 1997; Yasuda et al. 1997).  Nevertheless,
studies by
Chamaraux et al. (1980), Giraud (1986), Kenney \& Young (1986, 1989),
and Cayatte et al. (1990, 1994) show that the disk of NGC 4388 is
deficient in HI; it appears truncated at $\sim$ 10 kpc, well inside
the optical radius.  This suggests that NGC~4388 is experiencing the
effects of ram-pressure stripping by the densest portions of the
intracluster medium (ICM) in the Virgo cluster.  Distance
estimates using the Tully-Fisher relation suggest that NGC 4388
lies close to the mean distance of the M87 subcluster in the Virgo
cluster, i.e. near the core of the Virgo cluster (Mould, Aaronson, \&
Huchra 1980; Sandage \& Tammann 1984; Pierce \& Tully 1988; Yasuda et
al. 1997).  In the rest of this paper, we will assume that NGC~4388 is
indeed a member of the Virgo cluster and will adopt a distance of 16.7
Mpc (Yasuda et al. 1997), yielding a scale of 81 pc/$\arcsec$.

NGC~4388 was the first Seyfert galaxy discovered in the Virgo
cluster (Phillips \& Malin 1982). Nuclear activity has
been detected at nearly all wavelengths. The morphology of the nuclear
radio source suggests a collimated AGN-driven outflow (Stone
et al. 1988; Hummel \& Saikia 1991; Kukula et al. 1995; Falcke,
Wilson, \& Simpson 1998).  The nucleus hosts a strong hard X-ray
source with power-law photon index $\Gamma$ = 1.5, that is heavily absorbed by
a column density $N_H$ = 4 $\times$ 10$^{23}$ cm$^{-2}$ (Hanson et
al. 1990; Takano \& Koyama 1991; Lebrun et al. 1992; Iwasawa et
al. 1997). The large equivalent width of the iron K$\alpha$ line in
the ASCA spectrum of the nucleus (Iwasawa et al. 1997) and the detection
of broad recombination-line emission outside (Shields
\& Filippenko 1988, 1996) but not at the nucleus (e.g., Miller 1992;
Goodrich, Veilleux, \& Hill 1994; Veilleux, Goodrich, \& Hill 1997)
are both consistent with a powerful hidden AGN in NGC~4388.

At optical wavelengths, NGC 4388 has been known for some time to
present extended line emission (e.g., Ford et al. 1971; Sandage 1978;
Colina et al. 1987; Pogge 1988; Corbin, Baldwin, \& Wilson 1988). A
rich complex of ionized gas extends both along the disk of the
galaxy and up to 50$\arcsec$ (4 kpc) above that plane.
The extraplanar gas component has considerably higher
ionization than the disk gas, and appears roughly
distributed into two opposed radiation cones that emanate from the
nucleus (Pogge 1988; Falcke et al. 1998). Detailed spectroscopic
studies strongly suggest that this component is mainly ionized by
photons from the nuclear continuum (Pogge 1988; Colina 1992; Petitjean \&
Duret 1993).

Several origins of this extraplanar gas component have been suggested:
(1) it is tidal debris from a past interaction with another galaxy
(Pogge 1988); (2) supernovae have ejected it into the halo of the
galaxy (e.g., Corbin et al. 1988); (3) ram pressure from the ICM has
elevated some of it from the disk (Petit-Jean \& Durret 1993); (4) it
has been ejected from the nucleus by the mechanism that is powering
the radio jet (e.g., Corbin et al. 1988).

%(4) the gas constitutes
%coronal arches supported by the action of magnetic fields (e.g.,
%Wehrle \& Morris 1987) 

This ambiguity stems from the weak constraints on the
kinematics of the extraplanar gas that arise from the poor spatial
coverage of the spectra used in previous studies.
In this paper we use a Fabry-Perot spectrometer to
provide complete two-dimensional coverage of the kinematics of the
line-emitting region in NGC~4388. After a brief description of the
data analysis methods (\S 2), we present the
distribution and velocity field of the H$\alpha$ and [O~III]
$\lambda$5007 line-emitting gas derived from the data (\S 3). These
data allow us to split the line-emitting gas into two
distinct kinematic components, one associated with the galaxy disk
and the other associated with the extraplanar gas. In \S 4 we
derive dynamical parameters of the extraplanar gas
and use them to constrain the origin of this component. 
The important issue of ram-pressure stripping by the ICM is
addressed in the context of our results.  In this section, we
sketch a picture of NGC~4388 that accounts for the effects of
both the supersonic ICM and nuclear activity (Fig. 5). Our conclusions are
summarized in \S 5.

\section{Observations and Data Reduction}

The Hawaii Imaging Fabry-Perot Interferometer (HIFI; Bland \& Tully
1989) was attached at the Cassegrain focus of the University of Hawaii
2.2 m telescope at the Mauna Kea Observatory to obtain in photometric
conditions 54 seven-minute exposures of redshifted H$\alpha$ and 38 
eleven-minute exposures of redshifted [O~III] $\lambda$5007. The H$\alpha$
spectra were taken on 1990 May 31 and 1990 June 1 with a thick
epitaxial GEC CCD (5 $e^-$ rms readout), resulting in a pixel scale of
0$\farcs$66 pixel$^{-1}$. A finesse 70 etalon with free spectral
range 57 \AA~produced a velocity resolution of 40 km
s$^{-1}$. The [O~III] spectra were acquired on 1992 March 12 with a
blue-sensitive 1024 $\times$ 1024 Tektronix CCD and a finesse 60
etalon having a free spectral range of 61 \AA. The spatial scale and
spectral resolution of these data are 0$\farcs$85 pixel$^{-1}$ and
60 km s$^{-1}$, respectively.  Order separating filters with a flat-topped
transmission profile centered at 6614 \AA~(55 \AA~FWHM) and 5030
\AA~(36 \AA~FWHM) passed only one etalon order so true emission line
profiles of H$\alpha$ and [O~III] $\lambda$5007 could be
synthesized. Wavelength calibration exposures were made of diffuse Neon
(H$\alpha$ cube) and Mercury-Cadmium ([O~III] cube) sources.
A current-stabilized, flat continuum
source served to flatfield the images. Finally, the photometric
standard stars HD 94808, $\eta$ Hya, and $\theta$ Vir were observed
through the same setup as the object to provide an absolute flux
calibration. 

These data were reduced using the methods described in Bland \& Tully
(1989), Veilleux et al. (1994), and Bland-Hawthorn (1995). 
However, the H$\alpha$ data suffered from several cosmetic defects associated
with the aging GEC CCD.  While only a fixed bias offset needed to be
subtracted from the [O~III] data, row-by-row bias subtraction was
done on each H$\alpha$ frame to remove vertical striations of
the bias pattern. After interpolating over the
columns with poor charge transfer, the images of each night were
stacked in a cube in increasing order of wavelength. Each data set was
flatfielded using a similarly constructed whitelight cube. This
method produced good flat fields for the [O~III] spectra.
Large-scale illumination gradients in the sky background that persisted
in the corrected H$\alpha$ cube
were removed by dividing the spectra by a spatially smoothed
sky frame. This frame was obtained by summing all of the
frames in the data while interpolating through regions with
H$\alpha$ emission from the galaxy.  The photometric
accuracy resulting from this method is discussed in \S 3.1. The
rest of the reductions was the same for both H$\alpha$ and [O~III]
cubes: spatial registration of the
frames, cosmic ray removal, correction for air mass variations, sky
continuum and line subtraction, correction for the curvature of the
isovelocity surfaces (``phase calibration''), and flux
calibration. The OH sky line at 6604.13 \AA~(Osterbrock \& Martel
1992) was also used for the velocity calibration of the H$\alpha$ data.

For the sake of simplicity, the observed emission-line profiles were
approximated by simple Gaussian functions (regions where the
emission-line profiles differ from a simple Gaussian will be discussed
in \S 3.2).  The best fits were determined by the least-chi-squared
method on spectrally smoothed emission line profiles using a 1/4 --
1/2 -- 1/4 spectral filter (Hanning smoothing). Gaussian smoothing of
each image with $\sigma$ = 1.0 pixel improved the sensitivity to
fainter features. In the end, the centroid, intensity, and width were
successfully derived from $\sim$ 5,000 H$\alpha$ spectra and $\sim$
7,000 [O~III] $\lambda$5007 spectra.

\section{Empirical Results}

\subsection{Distribution and Excitation Properties of the Ionized Gas}

Figure 1 presents the distributions of H$\alpha$ and [O~III]
$\lambda$5007 line emission that we derived from the Gaussian fits; the
underlying starlight has been removed. The 1-$\sigma$ detection limit
of our data corresponds to a surface brightness of $\sim$ 7 and 3
$\times$ 10$^{-17}$ erg cm$^{-2}$ s$^{-1}$ arcsec$^{-2}$ for H$\alpha$
and [O~III] $\lambda$5007, respectively. 
The flux calibration of our data cubes was checked against published
long-slit spectrophotometry (e.g., Pogge 1988; and Petitjean \& Durret
1993). The H$\alpha$ and [O~III] fluxes agree with published ones to
within $\sim$ 25 -- 50\%. 
Note that the H$\alpha$ and
[O~III] fluxes were not corrected for Galactic reddening because 
extinction is negligible toward NGC~4388 [E(B--V)
= 0.028; Burstein \& Heiles 1978, 1984].  The patchy dust
lanes running roughly east-west in the optical continuum images of
this galaxy (e.g., Phillips \& Malin 1982) suggest highly variable
internal extinction which may affect the apparent distribution of the
ionized gas near the galactic plane.  No attempt was made to correct
for this effect. Dilution of the H$\alpha$ fluxes by underlying Balmer
absorption is negligible in NGC~4388. Moreover, the Balmer absorption
profile is dominated by pressure broadening in the atmospheres of
early-type stars, making this feature indistinguishable from a slowly
varying continuum over the narrow spectral bandpass of our instrument.  

The most prominent features in the H$\alpha$ image are the strings of
H~II regions in the shape of two `spiral arms' within the disk of the
galaxy.  Strong H$\alpha$ disk emission is detected out to $\sim$ 5
kpc east of the nucleus (its extent is limited by the velocity
coverage of our data cube).  The bright H~II region complexes in this
part of the disk each have L$_{{\rm H}\alpha}$ $\approx$ 10$^{38.5}$
-- 10$^{39.7}$ erg s$^{-1}$.  Emission near the detection limit of our
H$\alpha$ data is observed projected north of the plane of the disk
along PA = 20$\arcdeg$ -- 50$\arcdeg$ and 1 -- 4 kpc from the
nucleus. In contrast, the [O~III] map is dominated by nuclear line
emission.  Emission from the spiral arms and from gas above the galaxy
disk is also visible but considerably fainter. Its large projected
$\vert$z$\vert$ distance implies that the gas is well above the plane
of this highly inclined ($\vert i\vert$ = 78$\arcdeg$ $\pm$
1$\arcdeg$; cf. \S 3.2) galaxy. If this gas was in the disk, it would
lie $\sim$ 20 kpc from the nucleus which greatly exceeds the visible
radius of the galaxy ($R_{25}$ = 14 kpc from de Vaucouleurs et
al. 1991).

The differences between the H$\alpha$ and [O~III] line emission
distributions is particularly evident in Figure 2, which maps the
[O~III]/H$\alpha$ flux ratios.  This map fundamentally compares the
ionization level of the line-emitting gas. Energies of 35.5 eV and
54.9 eV are needed to produce O$^{++}$ and O$^{+++}$,
i.e. significantly larger than the 13.6 eV needed to ionize
hydrogen. Uncertainties on ratios range from 25\% in the bright H~II
regions in the disk to $>$50\% in the fainter high-$\vert$z$\vert$ gas
(recall that the extraplanar H$\alpha$ emission is near the detection
limit of the data). The low-ionization gas is found in the bright H~II
regions of the spiral arms. These H~II regions have [O~III]
$\lambda$5007/H$\alpha$ ratios of $\sim$0.1, typical of star-forming
regions with low ionization temperature ($\sim$ 40,000 K; Evans \&
Dopita 1985) or high metal abundance (O/H $\sim$ 3 $\times$
O/H$_\odot$; McCall, Rybski, \& Shields 1985).  The high-ionization
gas concentrates into two regions north and south of the nucleus. The
southern extension has a conical shape with apex near the nucleus.
Typical values for the [O~III] $\lambda$5007/H$\alpha$ ratio in the
nucleus and in the ionization cones are 1 to 5, consistent with
photoionization by the AGN (Pogge 1988; Colina 1992; Petitjean \&
Durret 1993).  This ionization map is qualitatively similar to
previously published [O~III] $\lambda$5007/(H$\alpha$ + [N~II]
$\lambda\lambda$6548, 6583) ratio maps obtained with narrow-band
interference filters (Pogge 1988; Corbin et al. 1988; Wilson \&
Tsvetanov 1994; Mulchaey, Wilson, \& Tsvetanov 1996; Falcke et
al. 1998). Note, however, that the present line ratio map was derived
directly from our spectra and is therefore not subject to the large
uncertainties in the line separation and continuum subtraction which
often affect line ratio maps derived from narrow-band images.

\subsection{Kinematics of the Extraplanar Material}

Figure 3 presents the barycentric velocity fields that are derived from the 
H$\alpha$ and [O~III] $\lambda$5007 spectra; uncertainties
range from $\sim$20 km s$^{-1}$ in the bright line-emitting regions to
$\ga$100 km s$^{-1}$ in the fainter areas. The overall velocity
field of the [O~III]-emitting gas resembles that of the
H$\alpha$-emitting gas. In both, the velocity field in the disk
is characterized by a large-scale east-west gradient indicative of
rotation in the plane of the disk. The rotation curve derived from our
data is highly consistent with earlier results (cf. Rubin, Kenney, \&
Young 1997 and references therein). Non-Gaussian H$\alpha$ and [O~III]
$\lambda$5007 profiles are found near the nucleus (R $\la$ 10
$\arcsec$), again confirming earlier results (Iye \& Ulrich 1986; Colina
et al. 1987; Ayani \& Iye 1989; Veilleux 1991; Rubin et al. 1997).

Figure 3 shows a kinematic dichotomy between the disk gas and
the extraplanar gas. To more objectively identify which portion of the
line-emitting gas does not partake in disk rotation, we have
constructed kinematic models to simulate the observed velocity field
in the disk. These models are described in more detail in VBC.  Because
the focus of the present paper is on the origin of the extraplanar gas
in NGC~4388, only the main results of our analysis of the
disk velocity field will be discussed here.

To minimize possible effects of the active nucleus, our analysis of
the disk component focused on the H$\alpha$ velocity field.  To
summarize our results, Figure 4 compares the observed H$\alpha$
velocity field with our best-fitting axisymmetric disk model and our
best-fitting disk model that includes bar streaming motions.  Compared
to the model with purely circular motions, the velocity residuals from
the model incorporting elliptical streaming are significantly smaller
throughout the disk. The residuals have a symmetric Gaussian
distribution with dispersion of only 10 km s$^{-1}$. This is an
excellent fit considering that hydrodynamic effects (e.g., strong
shocks along the bar; Athanassoula 1992; Piner, Stone, \& Teuben 1995)
are ignored in our simple kinematic model.  The parameters of the
model with elliptical streaming are listed in Table 1 and discussed in
more detail in VBC. The strong bar streaming motions prevented us from
deriving a rotation curve directly from the data; instead, we used a
smooth parametric function to model the rotation curve (cf. eqn. 3 in
VBC).  Note that the inclination of the disk, $i = -78\arcdeg$,
implies that the north rim of the disk is the near side (i.e. we are
looking at the galaxy from under the disk).

Our kinematic model suggests that most of the gas in the disk of
NGC~4388 partakes in bar-forced orbital motion around the
nucleus.  There are, however, two notable exceptions. The H$\alpha$ and
[O~III]-emitting gas with R $\ga$ 1.2 kpc and PA = 10$\arcdeg$ --
70$\arcdeg$ has velocities which appear blueshifted by 75 -- 250 km
s$^{-1}$ relative to systemic velocity. The location and velocity
field of this gas implies that it is elevated above the disk of
the galaxy.

No obvious kinematic trends with height above the disk or with
galactocentric radius are seen in the NE complex.  However, a
correlation appears to exist between the velocity field and the
surface brightness of the gas in the NE complex.  The most extreme
velocities in this gas complex appear in the bright cloud located
near the nucleus at R $\approx$ 2 kpc and PA $\approx$ 55$\arcdeg$ (v =
--~250 km s$^{-1}$; the `NE cloud' in the nomenclature of Pogge
1988). Double-peaked [O~III] profiles with velocity splitting $\sim$140 
km s$^{-1}$ are seen near the north-east edge of that cloud.  The much
fainter gas directly east of this NE cloud has velocities which
average 175 km s$^{-1}$ larger (--75 km s$^{-1}$ relative to systemic)
than that of the cloud. The trend between surface brightness and
blueshifted velocities is also seen in the cloud complex at R
$\approx$ 3.3 kpc and PA $\approx$ 35$\arcdeg$. Here, the brighter clouds have
velocities which are $\sim$ 60 -- 100 km s$^{-1}$ more blueshifted
than the diffuse gas surrounding them.

The bright [O~III] cloud south-west of the nucleus at R
$\approx$ 1.8 kpc and PA = 215$\arcdeg$ also has anomalous velocities
relative to the disk. This cloud is not detected at H$\alpha$,
implying [O~III]~$\lambda$5007/H$\alpha$ $\ga$ 5. Substructures
(shoulders, line splitting) of order $\sim$ 200 km s$^{-1}$ are
seen in the emission-line profiles on the north-east side of that cloud
(also detected by Corbin et al. 1988). A comparison with the modeled
disk velocity field indicates that this cloud has an average redshift
of $\sim$ 100 -- 150 km s$^{-1}$ relative to the underlying disk
gas.  Such supersonic motion through the disk ISM is highly
unlikely; therefore, this cloud is probably in front of the disk. Its
radial velocity relative to systemic is only $\sim$ 20 -- 30 km s$^{-1}$.

%region B of Corbin et al. ... velocity field can be fitted with 
%disk rotation???
Contrary to Petitjean \& Durret (1993), we see no general trend for
gas at large radii to have negative velocities relative to the galaxy
rotation. The probable origin of this discrepancy is the use of
different kinematic models to reproduce the observed velocity
field. As we demonstrated in \S 3.1, elliptical streaming is important
in the disk of NGC~4388 but this effect was not incorporated in the
axisymmetric models of Petitjean \& Durret (1993). We feel that the
complete two-dimensional coverage of our spectra and our use of
slightly more sophisticated models put our conclusions on a more solid
basis.

\section{Discussion}

\subsection{Derived Properties of the Extraplanar Material}

A clue to the origin of the extraplanar gas is provided by the
energetics of this material.  In this section, we estimate the
dynamical time scale, mass, and kinetic energy of the extraplanar gas
in NGC~4388.

An upper limit on the dynamical time scale of the extraplanar gas can
be estimated from its observed velocity and linear extent: $t_{\rm
dyn} \approx R/V \la 2 \times 10^7~(R/4~{\rm kpc})~(V_{\rm
obs}/200~{\rm km~s}^{-1})^{-1}~~~{\rm yr}$.  The ionized mass above
the disk follows from the [O~III] $\lambda$5007 flux.  Other
researchers have estimated the density and temperature of the
extraplanar gas.  Pogge used the [S~II] $\lambda$6731/$\lambda$6716
flux ratio to find densities in the range 100 -- 1,000
cm$^{-3}$. However, densities $<$50 cm$^{-3}$ are claimed by Colina
(1992) and Petitjean \& Durret (1993) for the NE complex.
%A lower limit on the density of the NE cloud may be derived
%from its [O~III] flux if we assume that it is spherical with a radius
%$\approx$ 1.5$\arcsec$ and a temperature $\sim$ 10$^4$ K; we find
%$N_e$(NE cloud) $\approx$ [1]/$\epsilon^{1/2}$ cm$^{-3}$ $\ga$ [1]
%cm$^{-3}$ where $\epsilon$ is the filling factor ($<$ 1).  
Estimates for the temperature range from $\sim$ 10,000 K (Petitjean \&
Durret 1993) to 15,000 K (Colina 1992).  In the following discussion,
we parametrize the mass in terms of the density and use an electron
temperature of 10$^4$ K.

The total [O~III] intensity of $\sim$ 2.5 $\times$ 10$^{-13}$ erg
s$^{-1}$ cm$^{-2}$ beyond R = 1.2 kpc and within PA = 10 --
70$\arcdeg$ yields an ionized mass in the NE complex of M(NE)
$\approx$ 4 $\times$ 10$^4$ X$^{-1}$ N$_{e,2}^{-1}$ M$_\odot$, where X
is the fraction of doubly ionized oxygen and N$_{e,2}$ is the electron
density in units of 100 cm$^{-3}$. In this calculation, we used a
solar mass fraction for oxygen and a five-level atom approximation to
estimate the emission coefficient (McCall 1984).  A lower limit on the
bulk kinetic energy of the north-east gas, $E_{\rm bulk}$(NE), may be
derived by summing the bulk kinetic energy over each pixel in this
region ($\Sigma_i$ 1/2 $\delta m_i$ $v_i^2$, where $\delta m_i$ and
$v_i$ are the mass and observed radial velocity at pixel $i$). We find
$E_{\rm bulk}$(NE) $\ga$ 6 $\times$ 10$^{51}$ X$^{-1}$ N$_{e,2}^{-1}$
ergs. The ``turbulent'' kinematic energy of this gas, $E_{\rm
turb}$(NE), may be derived from $\Sigma_i$ $\delta m_i$ $\sigma_i^2$,
where $\sigma_i$ = V$_{\rm FWHM}$/1.67; we get $E_{\rm turb}$(NE)
$\approx$ 1 $\times$ 10$^{52}$ X$^{-1}$ N$_{e,2}^{-1}$ ergs. Following
a similar procedure for the SW cloud, we find M(SW) $\approx$ 1,000
X$^{-1}$ N$_{e,2}^{-1}$ M$_\odot$, $E_{\rm bulk}$(SW) $\ga$ 6 $\times$
10$^{49}$ X$^{-1}$ N$_{e,2}^{-1}$ ergs and $E_{\rm turb}$(SW)
$\approx$ 3 $\times$ 10$^{50}$ X$^{-1}$ N$_{e,2}^{-1}$ ergs.  Using X
= 1 and N$_{e,2}$ = 0.1, the total mass and kinetic energy of the
extraplanar gas are thus $\sim$ 4 $\times$ 10$^5$ M$_\odot$ and
$E_{\rm kin}$ = $E_{\rm bulk}$ + $E_{\rm turb}$ $\ga$ 1 $\times$
10$^{53}$ ergs.  Using our estimate of the dynamical timescale, we
derive a lower limit on the time-averaged rate of kinetic energy
injected into the extraplanar gas: $dE_{\rm kin}/dt \simeq E_{\rm
kin}/t_{\rm dyn}~\ga 2 \times 10^{37}~N_{e,2}^{-1}~~~{\rm
erg~s}^{-1}$. We must emphasize that the energetics derived in this
section only take into account the optical component of the
extraplanar gas. Entrainment of the molecular gas from the disk could
significantly increase the energetics of the extraplanar material.

\subsection{Origin of the Extraplanar Material}

Our results can constrain the origin of the extraplanar gas
in NGC~4388. In this section, the predictions of each of the
four scenarios introduced in \S 1 are discussed and compared with our data.

\subsubsection{Tidal Debris from a Past Galaxy Interaction}

In analogy with the Magellanic Stream (Mathewson, Schwarz, and Murray
1977; Lin, Jones, \& Klemola 1995), the extraplanar material in
NGC~4388 may represent the tidal debris from a recent encounter with a
small gas-rich galaxy (Pogge 1988).  In this scenario, gas tidally
stripped from the dwarf galaxy appears when ionized by the radiation
cone of the active nucleus.  This scenario has been suggested by
Tsvetanov \& Walsh (1992) to explain the narrow ionized ``arcs'' in
NGC~5252. In this S0 galaxy with an AGN, the ionized arcs are
invisible outside sharp-edged conical sectors. However, the arcs have
kinematics (Morse et al.  1998) and space distribution that continue
into H I gas (Prieto \& Freudling 1996) on a ring inclined to the S0
disk that is biconically illuminated, hence ionized, by nuclear
photons.

Numerical stellar dynamical simulations of the accretion of a dwarf
galaxy by a large disk galaxy show that the satellite sinks in only
$\sim$ 1 Gyr and causes considerable damage to the disk's inner
regions (Walker, Mihos, \& Hernquist 1996). Satellites on bound polar
orbits experience severe orbital decay, settling into an orbit
coplanar with the disk after only one or two orbits (Walker et
al. 1996).  The well-behaved velocity field of the disk in NGC~4388
and the presence of gas well above the disk therefore tightly
constrains the epoch of the purported encounter, i.e. $\la$ 10$^8$
yrs.  In this scenario, the velocity field of the tidal debris is
expected to reflect the orbital motion of the original dwarf galaxy,
showing trends with radial distance from the nucleus, the center of
the potential dwell. These trends are absent in our
data. Moreover, it is difficult to explain in this scenario the
velocity -- surface brightness correlation and the steep velocity
gradients observed in the NE complex (\S 3.2).

\subsubsection{Interaction with the Intracluster Medium}

NGC~4388 is one of the most HI-deficient spiral galaxies in the Virgo
cluster (Chamaraux et al. 1980; Cayatte et al. 1990; 1994).  A
group III galaxy in the nomenclature of Cayatte et al. (1994), it has
a severely truncated HI disk with an HI-to-optical size ratio $D_{\rm
HI}/D_{\rm opt}$ = 0.7.  NGC~4388 is located very close to the center
of the Virgo cluster. Taking the position of the core to be the peak
position of the luminosity density, $\alpha$ = 12h 25.3m and $\delta$
= +13$\arcdeg$ 18$\arcmin$ and adopting 16.5 Mpc for the core distance
(Binggeli, Tammann, \& Sandage 1987; Yasuda et al. 1997), we find that
NGC~4388 is $<$300 kpc from the center of the Virgo
cluster. Using $V_{\rm Virgo}$ $\approx$ 1,000 km s$^{-1}$ for the
recession velocity of the cluster (Yasuda et 1997), NGC~4388 is
plunging edgewise at --1,500 km s$^{-1}$ into the core.  It
is therefore a prime candidate for ram pressure and turbulent viscous
stripping by the ICM.

Ram-pressure stripping has been proposed by Petitjean \& Durret (1993)
to explain the velocity field of the extraplanar material in NGC~4388.
In this scenario, the line-emitting gas in the NE complex may
represent (1) gas that has cooled and condensed out of the
ICM or (2) interstellar gas originally in
the disk of the galaxy which has been elevated by ram-pressure
exerted by the fast-moving ICM.  The first possibility appears unlikely
on the basis that the cooling time scale of the ICM (Rangarajan et al.
1995),
\begin{eqnarray}
t_{\rm cool} \approx 4 \times 10^6(T_{\rm ICM}/10^7~K)^{3/2}
 (n_{\rm ICM}~{\rm cm}^{-3})^{-1}~~{\rm yr},
\end{eqnarray}
exceeds the Hubble time for the conditions of the Virgo ICM
($T_{\rm ICM}$ = 2.5 keV = 2 $\times$ 10$^7$ K, $n_{\rm ICM}$
$\approx$ 10$^{-4}$ cm$^{-3}$; Koyama, Takano, \& Tawara 1991;
B\"ohringer et al. 1994).  We therefore focus on the second scenario.  
% but is that t$_{cool}$ different from Whittle et al. 1986's ?

We suppose that the extraplanar gas arises from interstellar
gas that has been ram-pressure stripped from the galactic disk. In
this scenario, the ram pressure exerted by the ICM will be largest on
the {\em receding} eastern edge of the disk so this is the most likely
source of extraplanar gas; this is consistent with the presence
of high-$\vert$z$\vert$ gas east of the nucleus.  The swept-out gas
will be heated conductively by the hot ICM; it is therefore important
to determine if this gas will survive long enough in the ICM
environment to be detected optically at the observed large heights above
the disk.  Here, we consider only the HI gas phase because the
molecular gas distribution in NGC~4388 appears normal (Kenney \& Young
1989).  Following Cowie \& McKee (1977), we calculate the evaporation
timescale for a cloud embedded in the ICM. In the galaxy halo,
classical thermal conduction (i.e. no ordered magnetic fields that
would reduce the conductivity; see Hummel \& Saikia 1991) should
provide a reasonable estimate for the evaporation time, which is given
by
\begin{eqnarray}
t_{\rm evap} = 1,000 n_c R^2_{\rm pc} (T_{\rm ICM}/10^7~K)^{-5/2}_{\rm
ICM} ({\rm ln}\Lambda/30)~{\rm yr}
\end{eqnarray}
%from Vikhlinin et al. 1997
where $n_c$ is the embedded cloud density, $R_{\rm pc}$ is its radius
in parsecs, $T_{\rm ICM}$ is the temperature of the surrounding hot
medium, and $\Lambda$ is the Coulomb logarithm. For typical Galactic
HI clouds ($n_c$ $\approx$ 1 cm$^{-3}$, R $\approx$ 10 pc) embedded in
a hot medium of temperature 2.5 keV (Koyama, Takano, \& Tawara 1991;
B\"ohringer et al. 1994), the evaporation time is $t_{\rm evap} \approx$
10$^4$ yrs. This is much shorter than the dynamical timescale $t_{\rm
dyn}$ $\approx$ 2 $\times$ 10$^7$ years derived in \S 4.1.

Shock waves generated during the sweeping may also contribute to
heating the interstellar clouds. The radial velocity $V_r$ = 1,500
km s$^{-1}$ of NGC~4388 relative to the ICM would produce
shock waves in the ICM (where the sound speed $c_{\rm ICM} \approx$ 0.1
T$_{\rm ICM}^{1/2}$ km s$^{-1}$ $\approx$ 500 km s$^{-1}$) and in the
interstellar gas of NGC~4388 ($c_{\rm HI}$ $<$ 3 km s$^{-1}$, assuming
HI with T $<$ 10$^3$ K). The shock velocity in the HI
cloud will be $v_s \approx (n_{\rm ICM}/n_c)^{1/2}~V_r$ $\approx$ 15
km s$^{-1}$ (e.g., McKee \& Cowie 1975; Sgro 1975) and the gas in the
cloud will be heated to T $\approx$ 14 $v_s^2$ $\approx$ 3,200 K in a
time $t_h$ $\sim$ 2R/$v_s$ $\approx$ 10$^6$ yrs $<<$ $t_{\rm dyn}$.
%Most of the energy will be consumed in heating the gas as the thermal
%energy density 5$nkT$/3 is nearly as high as the ram pressure $p$
%$\approx$ $\rho_{\rm ICM}$ $V^2$ $\approx$ 10$^{-11?}$ erg cm$^{-3}$.
%(kotanyi et al. 1983) 

Another serious argument against the ram-pressure hypothesis is based on
the kinematics of the extraplanar gas.  The equation of motion
for the HI clouds that are stripped by the ICM and somehow survive
evaporation/heating is
\begin{eqnarray}
M_c~g = C\rho_{\rm ICM}(V_{\rm ICM}-V_c)^2l_c^2, 
\end{eqnarray}
where $g$ is the cloud acceleration, $M_c$, $V_c$, and $l_c$, are the
mass, speed, and characteristic size of the interstellar clouds, and
$\rho_{\rm ICM}$ and $V_{\rm ICM}$ are the mass density and speed of
the ICM, respectively. The value of the constant C depends on the
details of the cloud model (e.g., C $\approx$ 9 for an isothermal
cloud model; DeYoung \& Axford 1969; Longair, Ryle, \& Scheuer 1973;
Blandford \& Konigl 1979a,b). The equation of motion can be simplified
if we make the reasonable assumption that $V_c$ $<<$ $V_{\rm ICM}$ in
the present situation and also assume that the size and mass of the
clouds are constant. (In reality the cloud expands because of the
decrease in ram pressure and its mass may also decrease because of
erosion caused by instabilities). With these assumptions, the
distance $x_{\rm acc}$ over which the cloud needs to be accelerated to
attain velocity $V_c$ is given by
\begin{eqnarray}
x_{\rm acc} = {V_c^2\over 2g} = \left({M_cV_c^2\over 2C\rho_{\rm
ICM}V_{\rm ICM}^2l_c^2}\right).
\end{eqnarray}
Using $V_c$ $\approx$ 200 km s$^{-1}$ for the velocity of the
extraplanar gas and $M_c$ $\approx$ 20 M$_\odot$ for the mass of
typical HI clouds ($n_c \approx 1$ cm$^{-3}$; $l_c$ $\approx$ 10 pc)
entrained in the ICM, we get $x_{\rm acc}$ of only $\sim$ 100 pc.

This simple model therefore predicts velocities for the extraplanar
material at $\sim$ 2 -- 4 kpc which considerably exceed those
observed.  Large projection effects where much of the velocity of the
entrained gas is in the plane of the sky are unlikely. The location
and very large velocity of NGC~4388 relative to the Virgo cluster
($V_r$ = 1,500 km s$^{-1}$ $>>$ $\sigma_{\rm Virgo}$ = 762 km $^{-1}$,
the Virgo cluster velocity dispersion; Rangarajan et al. 1995)
strongly suggests that NGC~4388 is on a radial orbit roughly aligned
along our line of sight (otherwise, NGC~4388 would not be bound to the
Virgo cluster; see also Giraud 1986).  The observed tendency for gas
at large heights above the disk to have velocities close to systemic
is opposite to that expected for ISM entrained in the ICM. (To first
order, we expect $V_c$ $\propto$ $\vert z\vert^{1/2}$ in our simple
model; eqn.~4). Nor can the brightness -- velocity correlation in the
NE complex and the redshifted velocity of the SW cloud be explained in
this scenario. We conclude that the extraplanar gas component is not
exposed to the putative, high velocity ICM.  We discuss this
surprising result in more detail in \S 4.3.

\subsubsection{Supernova-Driven Event}

Corbin et al. (1988) have suggested that part of the NE complex (more
specifically, the NE cloud) may have been elevated above the disk
by a recent supernova. This scenario deserves serious
consideration because evidence for intense star formation is seen
throughout the disk of NGC~4388 (Fig. 1).  Associated with these
star-forming complexes are clustered supernovae that deposit
significant kinetic energy into the surrounding ISM. The kinetic
energy from the stellar ejecta will thermalize rapidly to
create an expanding ``bubble'' of very hot (T $\approx$ 10$^7$ --
10$^8$ K) gas.  Following Mac Low \& McCray (1988) and Mac Low,
McCray, \& Norman (1989), the condition for this bubble to ``break
out'' of the disk is given by
\begin{eqnarray}
D \simeq 3 \times
10^{-42}~(dE/dt)~H_{kpc}^{-2}~P_7^{-3/2}~n_0^{1/2}~~>~100, 
\end{eqnarray}
where $dE/dt$ is the rate at which energy is injected in the disk,
$H_{kpc}$ is the scale height of the galactic disk in kpc, $n_0$ is
the ambient density in cm$^{-3}$, and $P_7$ is the ambient pressure,
$P/k$, in units of 10$^7$ K cm$^{-3}$.  In analogy with our analysis
of the disk of NGC~3079 (Veilleux, Bland-Hawthorn, \& Cecil 1995), we
assume that the multi-phase nature of the ISM in NGC~4388 can be
simplified by using the following average disk parameters: $H_{kpc}$ =
1.0, $n_0$ = 1.0, and $P_7$ = 10$^{-3}$.  Under these circumstances,
break-out occurs if $dE/dt$ $\ga$ 1 $\times$ $10^{39}$ erg
s$^{-1}$. Note that this analysis neglects cushioning by the ambient magnetic
field (cf. Tomisaka 1990; Ferri\`ere, Mac Low, \& Zweibel 1991; Slavin
\& Cox 1992; Mineshige, Shibata, \& Shapiro 1993). We have therefore
probably underestimated the energy injection rate required for break-out.

Are any of the star-forming regions in the disk of NGC~4388 sufficiently
powerful to eject the NE cloud out of the disk? Based on the projected
location of the NE cloud, the best candidate HII region complex for
this event lies at $R \approx$ 1.9 kpc east of the nucleus. This complex
has L(H$\alpha$) $\approx$ 2 $\times$ 10$^{39}$ erg s$^{-1}$. 
Using equation (4) of Veilleux et al. (1995) to translate
this H$\alpha$ luminosity into a time-averaged rate of energy
injection from stellar winds and supernovae, we obtain $dE/dt$ $\approx$ 3
$\times$ 10$^{39}$ erg s$^{-1}$, barely sufficient for
break-out. Note that our analysis has not accounted for 
the kinematic energy in the NE complex itself:
$E_{\rm kin}$(NE cloud) = $E_{\rm
bulk}$(NE cloud) + $E_{\rm turb}$(NE cloud) $\ga$ 5 $\times$ 10$^{51}$
ergs, is quite substantial -- the equivalent to the energy released
from $\ga$ 5 Type II SNe.  

% [This would imply a SN rate which is unusually high in a Sb galaxy 
% like NGC~4388... (ref)]

Our dynamical analysis suggests that it is difficult to explain the NE
cloud as a result of star formation.  But there are two
important assumptions.  First, we neglected dust obscuration by the
edge-on disk when deriving the energy injection rate from the
H$\alpha$ flux of the eastern HII region.  Extinction estimates
towards the nucleus (A$_V$ $\approx$ 1 -- 2 mags; Phillips \& Malin
1982; Storchi-Bergmann et al. 1990; Osterbrock et al. 1992; Petitjean
\& Durret 1993) suggest that the correction may be substantial.
Second, it is unclear if the {\em current} energy injection rate
from this complex is relevant to an event that presumably took place
$z$(NE)/$V$(NE) $\approx$ 10$^7$ years ago. 
% the energy injection rate from a starbursting complex is known to 
% decrease by several orders of magnitude over that period of time; 
% e.g., Leitherer \& Heckman 1995).  
Evidently, dynamical arguments alone fail to reject the supernova hypothesis.

Nonetheless, we consider this hypothesis unlikely for three
further reasons: (1) Perhaps most damaging is the
morphology of the NE complex.  It does not resemble the
filaments/loops produced by intense star formation in the disks of
galaxies (e.g., Dettmar 1992; Veilleux et al. 1995).  A bridge of
blueshifted gas between the NE cloud and the nucleus is detected in
our [O~III] data. This is strongly suggestive of mass transfer from
the nucleus to the NE cloud rather than from the disk (see next
section).  (2) The overall blueshifted velocity field of the NE
complex is difficult to explain in this scenario.  Gas ejected by
SN-driven events is expected to have a strong velocity component
perpendicular to the plane of the disk (e.g., Norman \& Ikeuchi 1989).
Because $i$(NGC~4388) = --78$\arcdeg$, we expect that most of this gas
will be traveling {\em away} from us, not toward us. In addition,
the SN-event hypothesis cannot explain the presence of the SW
cloud. (3) The relatively long orbital period in the disk
at $R \approx$ 2 kpc (2 $\pi$ $R/V_{\rm orb}$ $\approx$ 10$^8$ years
$>$ $t_{\rm dyn}$) implies that morphological (e.g., filaments,
bubbles) or kinematical (e.g., non-rotational motion) disturbances
caused by a supernova should still be visible in the disk.  None are
detected in our data.

\subsubsection{Nuclear Outflow}

NGC~4388 has an elongated, jet-like radio
structure of nonthermal origin which extends $\sim$ 200 pc
south of the nucleus (e.g., Stone et al. 1988; Carral, Turner, \& Ho
1990; Hummel \& Saikia 1991; Kukula et al. 1995; Falcke et al. 1998).
This suggests a collimated nuclear outflow. This radio morphology
resembles that of several other Seyfert galaxies where the ejected
radio plasma is clearly interacting with the surrounding ISM,
producing elongated optical narrow-line regions where the gas is
compressed and entrained in the outflow, but is probably kept ionized by the
nuclear continuum (e.g., Whittle et al. 1988).

Is this phenomenon occurring in NGC~4388?  The orientation of the
putative nuclear outflow in NGC~4388 does not favor strong interaction
between radio plasma and ISM in the galaxy disk: the nuclear radio
source (i.e. the nuclear component and ``blob A'' in the nomenclature
of Hummel \& Saikia 1991) is elongated along P.A. $\approx$
23$\arcdeg$, i.e. close to the minor axis of the galaxy disk. A
diffuse bubble-like structure extends $\sim$ 1.5 kpc above the nucleus
due north (P.A. $\approx$ 5$\arcdeg$) of the galaxy disk.  This
strongly suggests a nuclear outflow which has burst out of the galaxy
plane.  The existence in Figure 3 of a continuous bridge of
approaching (blueshifted) gas between the nucleus and the NE complex
also supports this picture. The presence of a redshifted cloud
south-west of the nucleus, i.e. diametrically opposite to the
blueshifted NE complex suggests that the outflow is bipolar.  The mass
and kinetic energy involved in the extraplanar gas (cf. \S 4.1) are
similar to those involved in galactic-scale outflows in nearby
starburst and Seyfert galaxies (e.g., Cecil, Bland, \& Tully 1990;
Heckman, Armus, \& Miley 1990; Veilleux et al. 1994; Veilleux \&
Bland-Hawthorn 1997; Shopbell \& Bland-Hawthorn 1998). Line splitting
near the NE and SW clouds may reflect regions of high dissipation of
kinetic energy caused by the interaction/deflection of the outflowing
gas with ambient halo gas (e.g., Gallimore et al. 1996a, 1996b). The
blueshifted velocities of the extraplanar gas above the north (near)
side of the galaxy and redshifted velocities below the south (far)
side suggests that the radio jet tilts relative to the normal to the
disk by at least 90$\arcdeg$ -- $\vert i\vert$ $\approx$ 12$\arcdeg$.

One difficulty with the outflow scenario is the lack of correlation
between radio features in NGC~4388 and the extraplanar optical
gas.  Standard plasmon-driven bowshock models predict a
one-to-one correspondence between the radio and optical structures
(e.g, Pedlar, Dyson, \& Unger 1985; Taylor et al. 1989).  While the
lack of obvious optical counterpart to radio blob A can be attributed
to dust obscuration by the disk, the origin of the large offset
between the NE line-emitting complex in NGC~4388 and the northern
radio bubble is harder to explain.  The NE complex is
three times larger than the northern radio bubble, and is
misaligned from the axis of the radio bubble by $\sim$ 30$\arcdeg$.
%standard plasmon models consider the effects
%of an expanding radio-emitting component on a medium of constant
%density. This is clearly not the case in NGC~4388, where the radio
%plasma has burst through the disk and is now propagating through the
%halo of the galaxy. The authors are not aware of published simulations
%of relativistic plasmons that take these effects into account. We
%therefore have to rely on observations.  

However, morphological differences between radio and optical features
are common among edge-on galaxies with confirmed large-scale galactic
outflows (e.g., Colbert et al. 1996a, 1996b, 1998; Lehnert \& Heckman
1996).  In NGC~3079, for instance, most of the radio emission lies
outside the kpc-scale optical nuclear superbubble (Veilleux et
al. 1994), and X-shaped optical filaments with no obvious radio
counterparts rise more than 4 kpc above the disk plane (Heckman,
Armus, \& Miley 1990; Veilleux et al. 1995).  Several processes may
explain the lack of correlation between radio and optical emission in
NGC~4388. The lighter radio plasma is more susceptible to external
forces than the denser optical filaments. Buoyancy due to density
differences between the radio cloud and ambient gas in the halo and
the galactic gravitational gradient (parallel to the minor axis of the
disk) have been suggested to explain the bending in the radio
structure (Stone et al. 1988).  Here the net buoyancy force on a cloud
of density $\rho_c$ immersed in a region of density $\rho_0$ is
$\propto$ $\rho_0 - \rho_c$.  Buoyancy will have no direct effects on
the denser optical line-emitting gas, thereby kinematically decoupling
this gas from the radio plasma.  Balancing the buoyancy on the radio
cloud against the ram pressure from the ambient halo gas and assuming
$\rho_c$ $<<$ $\rho_0$, Stone et al. (1988) derived a terminal
velocity for the radio cloud of 10 -- 35 km s$^{-1}$, i.e. much
smaller than the apparent velocity of the {\em line-emitting}
gas. Interestingly, the time needed to reach the current radio extent
at this speed is similar to the dynamical timescale for the
line-emitting gas ($t_{\rm dyn}$ $\approx$ 2 $\times$ 10$^7$ yrs).
The lighter radio plasma will also be more subject to refractive
bending by density gradients in the halo gas than the dense
line-emitting gas (Henriksen, Vall\'ee, \& Bridle 1981; Smith \&
Norman 1981; Fiedler \& Henriksen 1984).  A steep vertical density
gradient in the ambient gas will force the radio plasma to follow a
path close to the minor axis of the galaxy, as seen in the radio
data. Finally, magnetic forces may also act to confine the lighter
radio plasma (e.g., Wehrle \& Morris 1987) without substantially
affecting the kinematics of the more massive optical
filaments. Current radio data do not constrain the strength and
orientation of the magnetic field in the extraplanar gas component.

%The optical and radio features may reflect outflow events that took 
%place at different epochs.  

Assuming the morphological differences between the radio and optical
emission are indeed due to a kinematic decoupling between these two gas
components, one question still remains: is the radio jet in the nucleus of
NGC~4388 powerful enough to accelerate the gas in the NE complex to
its current velocity?  To answer this question we assume that the
acceleration phase of the line-emitting gas occurred near the
nucleus when the radio plasma was still coupled to the line-emitting
gas. The line-emitting clouds, initially at rest, were
accelerated downstream by the jet. Under these assumptions, the
equation of motion of the line-emitting clouds at this early epoch is
similar to that of clouds entrained in the ICM (eqn.~3).  The
distance $x_{\rm acc}$ over which the cloud must accelerate to
reach velocity $V_c$ is given by equation (4) after replacing
$\rho_{\rm ICM}$ and $V_{\rm ICM}$ by $\rho_j$ and $V_j$, the mass
density and velocity of the jet gas.  The value of $x_{acc}$ can
be estimated if we use the fact that $L_r = \epsilon
N_c\rho_jV_j^3l_c^2$/2 where $L_r$ is the luminosity of the radio jet,
$N_c$ is the number of clouds entrained by this jet, and $\epsilon$ is
the fraction of the total power of the jet which is converted into
radio emission (Wilson 1981):
\begin{eqnarray}
x_{acc} &=& {\epsilon~M_{tot}V_c^2V_j\over 4~C~L_r}, \nonumber \\
                     &=& \left(2~\epsilon\over
                     10^{-2}\right)\left(M_{tot}\over 10^6
                     M_{\odot}\right)\left(V_c\over 200~{\rm
                     km~s}^{-1}\right)^2\left(V_j\over 10^4~{\rm
                     km~s}^{-1}\right)\left(10\over
                     C\right)\left(10^{37}~{\rm erg~s}^{-1}\over
                     L_r\right) {\rm kpc}. 
\end{eqnarray}
Here, $M_{tot} = N_c~M_c \simeq 4 \times 10^5 ~M_\odot$ is the total
mass of line-emitting clouds entrained in the radio jet (cf. \S 4.1).
As a measure of the jet radio luminosity, we use the sum of the radio
luminosity in ``blob A'' and the ``elongated feature'' (Hummel \&
Saikia 1991; $L_r \approx 7 \times 10^{37}$ erg s$^{-1}$ for our
adopted distance of 16.7 Mpc).  A reasonable estimate for $x_{acc}$ is
therefore $\sim$ 100 pc, implying that the acceleration zone of the
line-emitting clouds is well within the radial position of the NE
cloud (R $\approx$ 2 kpc). Entrainment with the radio jet can
therefore produce the observed velocities of 200 -- 300 km
s$^{-1}$.  While the previous derivation neglects projection
effects and ram pressure by the ambient halo gas (both of which will
tend to increase the required acceleration distance, $x_{\rm acc}$),
we feel that the kinematic energy in the NE complex can be explained
through this nuclear outflow scenario. From equation (3), ram-pressure
deceleration by the ambient halo gas will be stronger for small clouds
(large l$_c^2$/$M_c$) than for large clouds, and may therefore explain
the larger blueshift of the brighter (more massive) clouds in the NE
complex.

\subsection{Nature of the Interaction with the Virgo Intracluster Medium}

The disk of NGC~4388 is oriented ($i$ = --78$\arcdeg$) such that the
extraplanar gas to the north-east of the galaxy should feel the
full force of the ICM `wind', while the high-$\vert$z$\vert$ gas to the
south may be somewhat `shielded' by the galaxy, lying in the
%[turbulent?] 
wake of the flow.  Simulations of radio jet propagation in a moving
ICM (e.g., O'Donohue, Eilek, \& Owens 1993; Loken et al. 1995) show
that the ram pressure from the ICM should have profound effects on the
morphology and kinematics of the outflowing gas.  The lack of
evidence for entrainment of the north-east extraplanar gas in the
Virgo ICM is therefore puzzling.  In this section we discuss in more
detail the nature of the interaction between NGC~4388 and the Virgo
ICM, and estimate the effects of this interaction on the kinematics of
the extraplanar material.

We first re-examine the rationale for ICM--ISM interaction in
NGC~4388.  Ram-pressure sweeping depends on the local pressure of the
intracluster medium, the relative motion of a galaxy relative to
the intracluster medium and the local surface density within the
galaxy.  Ram-pressure stripping of the gaseous phase of a galaxy will
occur if the ram pressure from the ICM exceeds the gravitational
restoring force in the galaxy (Gunn \& Gott 1972), i.e.  
\begin{eqnarray}
\rho_{\rm ICM}V_\bot^2 > 2 \pi G \sigma_{\rm tot} \sigma_{\rm gas}
\end{eqnarray}
where $\rho_{\rm ICM}$ is the density of the intracluster medium,
$V_\bot$ is the relative velocity of the intracluster medium
perpendicular to the galaxy plane, $\sigma_{\rm tot}$ is the galaxy
total surface density and $\sigma_{\rm gas}$ = $\sigma_{\rm HI}$ +
$\sigma_{{\rm H}_2}$ is the gas surface density.  Using $V_\bot$ =
1,200 km s$^{-1}$, $\rho_{\rm ICM}$ = 1.6 $\times$ 10$^{-4}$ cm$^{-3}$
(derived from a general model fit to the X-ray surface brigthness
profile; Takano et al. 1989, and references therein), $\sigma_{\rm
gas}$ = $\sigma_{\rm HI}$ = 10$^{20}$ cm$^{-2}$, and $\sigma_{\rm
tot}$ = 157 M$_\odot$ pc$^{-2}$ (calculated using a Brandt model
potential with n = 1.3 at the HI radius where $\sigma_{\rm gas}$ =
10$^{20}$ cm$^{-2}$), Cayatte et al. (1994; their Table 3) have argued
that condition (7) is met beyond R $\approx$ 6 kpc in NGC~4388. But
the large value of $V_\bot$ used in their derivation and the edge-on
orientation of NGC~4388 would imply that NGC~4388 has a large velocity
transverse to our line of sight, $V_t$ = ($V_\bot$ -- $V_r$~cos~$\vert
i\vert$)/sin~$\vert i\vert$ $\approx$ 900 km s$^{-1}$. Given the mass
of the Virgo cluster, $M_{\rm Virgo} \sim 5 - 9 \times 10^{14}
M_\odot$ (e.g., Tully \& Shaya 1984; Giraud 1986), the total velocity
of NGC~4388, [$V_t^2$ + $V_r^2$]$^{1/2}$ $\approx$ 1,800 km s$^{-1}$,
would imply that NGC~4388 is unbound to the Virgo cluster. We consider
this unlikely based on the HI deficiency of NGC~4388 and its proximity
to the center of the cluster.  More likely, NGC~4388 is in a minimum
potential energy part of its orbit and most of its velocity is along
our line of sight. In this case, $V_\bot$ $\approx$ $V_r$~cos~$\vert
i\vert$ $\approx$ 300 km s$^{-1}$, the ram pressure is 1/16th that
used by Cayatte et al. (1994), and stripping occurs only in the
outskirts ($R$ $>$ 10 kpc) of NGC~4388.

However, as noted by Nulsen (1982), ram pressure stripping is not the
only means by which the ICM can strip a galaxy. Momentum
transfer between the hot ICM and the galaxy ISM \textit{via} the
Kelvin-Helmholtz instability can increase stripping rates
under some circumstances.  While the effectiveness of the ram pressure
is extremely sensitive to orientation, the rate of mass loss from such
turbulent viscous stripping is approximately
\begin{eqnarray}
\dot{M}_{\rm visc} = \pi R_{\rm HI}^2 \rho_{\rm ICM} V \approx \pi R_{\rm HI}^2 \rho_{\rm ICM} V_r,
\end{eqnarray}
where $R_{\rm HI}$ is the radius of the HI disk and $V$ is the
velocity of NGC~4388 relative to the cluster.  Using the current
extent of the HI disk of NGC~4388, Cayatte et al. (1994) estimate that
$\dot{M}_{\rm visc}$(NGC~4388) $\approx$ 1 -- 6 M$_\odot$ yr$^{-1}$,
implying that it will lose virtually all of its atomic content
($M_{\rm HI}$ = 4 $\times$ 10$^8$ M$_\odot$ for a distance of 16.7
Mpc; Kenney \& Young 1989) during a crossing time $t_{\rm crossing}$
$\simeq$ 3 $\times$ 10$^9$ yrs through the cluster core. This suggests
that NGC~4388 is on its first passage through the cluster core.

So far we have assumed that the ISM of NGC~4388 is experiencing
the full impact of the high-velocity ICM. In reality the ICM `wind'
will be disturbed by the presence of the galaxy. In particular, because
the sound speed in the cluster gas is $c_s \approx$ 500 km s$^{-1}$ and
the radial velocity of NGC~4388 relative to the Virgo cluster core
is $V_r$ $\approx$ 1,500 km s$^{-1}$, the interaction of
the ISM of NGC~4388 with the cluster gas will be supersonic with
Mach number $M$ = $V/c_s$ $\approx$ $V_r/c_s$ $\approx$ 3.
In this situation, a Mach cone (bow shock) forms near
the leading edge of NGC~4388. The opening half-angle of the Mach cone
is $\phi$ = 2 tan$^{-1}$ ($1/M$) $\approx$ 40$\arcdeg$ (note the factor
of 2 missing in the expression used by Rangarajan et al. 1995). 
%[The projected opening half-angle projected on the sky $\phi^\prime$ is
%similarly tan($\phi^\prime$/2) = $c_s/V_{\rm perp}$, where $V_{\rm
%perp}$ is the velocity perperdicular to our line of sight. ] The
%geometry of the NGC~4388 - ICM interaction is illustrated in Figure 5. 
The ICM will be shocked at the Mach cone.  The
Rankine-Hugoniot jump conditions for adiabatic shocks ($\gamma$ =
5/3) yield a post-shock velocity $V_{ps}$ $\approx$ $V_{\rm ICM}$/3
= 500 km s$^{-1}$, a post-shock density $n_{ps}$ $\approx$ 3 $n_{\rm
ICM}$ = 3 $\times$ 10$^{-4}$ cm$^{-3}$, and a post-shock temperature
$T_{ps}$ = 3.7 $T_{\rm ICM}$ = 7.4 $\times$ 10$^7$ K. From equation
(1), the cooling time for the shocked ICM is $\sim$ 3 $\times$
10$^{11}$ yrs, thereby justifying our assumption of an adiabatic
shock. Given the rather wide opening angle of the Mach cone, the
extraplanar gas observed in NGC~4388 will be affected by the
shocked -- rather than the pre-shock -- ICM (Fig. 5).

The turbulent viscous stripping rate due to the shocked ICM will be
the same as that calculated for the pre-shock ICM (because $\rho_{\rm
ICM} V_{\rm ICM}$ = $\rho_{ps} V_{ps}$; cf. eqn.~8), but the ram
pressure on the extraplanar gas will be substantially
reduced. Assuming the shocked ICM flow lines roughly follow the
contours of NGC~4388 and taking $V_{\rm outflow}$ = 250 km s$^{-1}$ as
the velocity component of the outflowing NE cloud along the plane of
the galaxy, the ram pressure from the shocked ICM on the extraplanar gas
will be reduced by a factor $\rho_{\rm ICM}/\rho_{ps}$
$\times$ [($V_{\rm ICM} - V_{\rm outflow}$)/($V_{ps} - V_{\rm
outflow}$)]$^2$ $\approx$ 8.3 relative to that from the pre-shock ICM.
Under these conditions the acceleration distance, $x_{\rm acc}$, required
to increase the approaching velocity of the gas in the NE cloud
from 250 km s$^{-1}$ to 350 km s$^{-1}$ is approximately
\begin{eqnarray}
x_{\rm acc} = {(V_{\rm final}^2 - V_{\rm initial}^2)\over 2<g>} =
\left({M_c(V_{\rm final}^2 - V_{\rm initial}^2)\over 2C\rho_{ps}<(V_{ps} -
V_{\rm outflow})^2>l_c^2}\right),
\end{eqnarray}
 where $V_{\rm final}$ = 350 km s$^{-1}$, $V_{\rm initial}$ = 250 km
s$^{-1}$, and $<g>$ and $<(V_{ps} - V_{\rm outflow})^2>$ are time-averaged
quantities.  Using $M_c$ $\approx$ 20 M$_\odot$ and $l_c$ $\approx$ 10
pc for typical extraplanar line-emitting clouds [i.e., $n_c$ $\approx$
1 cm$^{-3}$ and $l_c$ $\approx$ $l_c$(HI cloud)], we get $x_{\rm acc}$
$\approx$ 6 kpc. 
%is this a lower limit because $M/l_c^2$ for the NE cloud as a whole 
% is several 10$^4$/100$^2$ = several instead of 0.2? but looking at the 
% more diffuse cloud we have several 10$^4$/1000$^2$ $<$ 0.1 so it may 
% not be too bad of an estimate. 
This is significantly larger than the 2 -- 3 kpc separation between
the NE cloud
and the rest of the NE complex at higher $\vert$z$\vert$. The implication
is that the ram pressure exerted by the shocked ICM has only a
small effect on the kinematics of the outflowing extraplanar material.

%It is interesting to note that the extraplanar gas beyond 4 kpc will
%be outside the Mach cone and will therefore feel the full force of the
%ICM wind; this may be the reason why no gas is seen beyond 4 kpc.

We conclude this section on a more speculative note by suggesting that
the impact of the shocked ICM on the galaxy and the extraplanar gas
component may also be cushioned by a gaseous halo in hydrostatic
equilibrium around NGC~4388.  As we discussed in \S 4.2.4, a static
gaseous halo in NGC~4388 would also help to explain possible shock
structures in the NE and SW clouds and the curvature of the radio
jet. Our own Galaxy is known to be surrounded by a hot ($\sim$ 10$^6$
K) highly ionized corona extending out to several kpc from the
Galactic plane (e.g., Spitzer 1990; Sembach \& Savage 1992; Wang
1992).  Quasar absorption-line studies (e.g., Lanzetta et al. 1995)
and deep X-ray images of spiral and elliptical galaxies (e.g.,
Fabbiano 1988; Heckman et al. 1990; Dahlem et al. 1996) show similar
gaseous halos.  The main question is whether such halos survive long
enough in the Virgo cluster environment.  Current soft X-ray images of
NGC~4388 (e.g., Matt et al. 1994; Angelo Antonelli, Matt, \& Piro
1997; Iwasawa et al. 1997; Colbert et al. 1998) show soft X-ray
emission of thermal origin, but the exact morphology of the X-ray gas
is ill-defined because of the strong X-ray background from the Virgo
ICM and the distorted halo in nearby M86 (Forman et al. 1979; White et
al. 1991; Rangarajan et al. 1995). The very existence of this halo in
M86, a giant E/S0 galaxy which is located close to the Virgo cluster
core (15.7 Mpc according to Jacoby et al. 1990 or 18.4 Mpc according
to Tonry et al. 1990) and is moving at a velocity $\sim$ --1,300 km
s$^{-1}$ relative to the cluster ($V_r$ = --227 km~s$^{-1}$; Jacoby et
al. 1990), gives us some confidence that at least some galaxies in
Virgo [albeit those more massive than NGC~4388: M(M86) $\approx$
10$^{12}$ M$_\odot$; e.g., Forman et al. 1979] can keep an X-ray halo
while experiencing virtually the same ram pressure as that on
NGC~4388.

\section{Conclusions}

The complete two-dimensional coverage of our Fabry-Perot spectra on
the line-emitting gas in NGC~4388 has allowed us to separate
kinematically the extraplanar and disk gas components. We find that
the velocity field of the high-$\vert$z$\vert$ gas is best described
by a bipolar outflow rather than ram-pressure stripping by the
fast-moving ICM in the Virgo cluster. We argue for a Mach cone with
opening angle $\sim$ 80$\arcdeg$ to explain these results.
Unfortunately, this Mach cone will be difficult to detect, even with
AXAF and XMM, because it is predicted to be roughly face-on and the
strong X-ray background from nearby galaxies and the ICM in the Virgo
cluster will dominate the emission.  Nevertheless, high-resolution
X-ray spectroscopy would be useful to constrain the velocity field
(and metallicity) of the X-ray gas and to help establish if the hot
gas near NGC~4388 is in hydrostatic equilibrium in the galaxy
potential or instead represents fast-moving, shocked ICM. Detailed
investigations of bow shocks in galaxy clusters have the potential to
help us determine the space velocities of galaxy members with respect
to the ICM.

From a theoretical viewpoint there is a need to explore the effects of
a fast-moving ICM as it interacts with a gas-rich galaxy.  Few such
models have been published (e.g., Kritsuk 1984; Tosa 1994), and none
considers the multi-phase nature of the ISM and the effects that this
interaction may have on star formation (e.g., through compression;
Kenney \& Young 1989; Cayatte et al. 1994). The ICM `wind' will also
affect the morphology of the nuclear ejecta from starburst galaxies
and AGN (e.g., wide-angle tailed radio sources in cluster environment;
Loken et al. 1995). Outflowing gas will stretch in the direction of
the ICM motion, thereby increasing the effective gaseous cross-section
of these galaxies and perhaps contributing significantly to the
population of C~IV absorption lines systems in quasars (e.g., Norman
et al. 1996).

\clearpage

\acknowledgments

We would like to thank Jim Stone for helpful discussions. SV is
grateful for support of this research by a Cottrell Scholarship
awarded by the Research Corporation, NASA/LTSA grant NAG 56547, and
Hubble fellowship HF-1039.01-92A awarded by the Space Telescope
Science Institute which is operated by the AURA, Inc. for NASA under
contract No. NAS5--26555.  JBH acknowledges partial support from the
Fullam award of the Dudley Observatory.

\clearpage

\begin{table*}
\caption{Kinematical Parameters of NGC~4388$^{(a)}$}
\begin{center}
\begin{tabular}{lll}
\tableline
\tableline
\noalign{\vskip 7.5 pt}
Component &Parameter & NGC~4388\\
\tableline
\noalign{\vskip 7.5 pt}
Disk &$V_{\rm sys}$  & 2,525 km s$^{-1}$\\
     & $V_{\rm max}$ (deprojected) &240 km s$^{-1}$  \\
     &P.A. (major axis)    & 90$\arcdeg$ \\
     &Inclination $i$ & --78$\arcdeg$ \\
Bar  &P.A. (intrinsic to disk) & 135$\arcdeg$\\
     &Inner radius $R_b$ & 1.5 kpc\\
     &Orbit eccentricity $R \le R_b$& 0.3\\
%     &Angular velocity $\Omega_b$& 120 km s$^{-1}$ kpc$^{-1}$\\
     &Outer radius $R_d$ &5.1 kpc\\
\noalign{\vskip 7.5 pt}
\tableline
\end{tabular}
\end{center}

\tablenotetext{^{(a)}}{Refer to Veilleux, Bland-Hawthorn, \& Cecil
(1999) for a detailed discussion of these parameters. }

\end{table*}

\clearpage

\begin{figure}
%\epsscale{0.5}
%\plotone{fig1_v4.ps}
\caption{ The distributions of the line emission in NGC 4388. (top)
H$\alpha$; (bottom) [O~III] $\lambda$5007.  The optical fluxes were
determined from Gaussian fitting of the emission-line profiles. North
is at the top and East to the left. The spatial scale, indicated by a
horizontal bar at the bottom of the image, corresponds to 12$\arcsec$,
or 1 kpc for the adopted distance of 16.7 Mpc. The optical continuum
nucleus is indicated in each panel by a cross. The flux scale is
logarithmic and is expressed in units of 10$^{-16}$ erg s$^{-1}$
cm$^{-2}$ arcsec$^{-2}$.  
%The flux scale of the H$\alpha$ map ranges
%from 2 $\times 10^{-14}$ erg s$^{-1}$ cm$^{-2} $arcsec$^{-2}$ in the
%nucleus to $\sim$ 7$\times 10^{-17}$ erg s$^{-1}$ cm$^{-2}$ arcsec$^{-2}$,
%while the scale of the [O~III] flux map ranges from 5 $\times
%10^{-14}$ erg s$^{-1}$ cm$^{-2} $arcsec$^{-2}$ in the nucleus to $\sim$ 3
%$\times 10^{-17}$ erg s$^{-1}$ cm$^{-2}$ arcsec$^{-2}$.  
}
\end{figure}

\begin{figure}
%\epsscale{0.5}
%\plotone{fig2.ps}
\caption{ [O~III] $\lambda$5007/H$\alpha$ flux ratio map of the
line-emitting gas in NGC 4388. Same spatial scale and orientation as
Fig. 1.  The optical continuum nucleus is indicated by a cross. Values
for this ratio range from $\sim$ 0.1 in the disk to $\ga$ 5 in the
extraplanar material. Regions with [O~III]~$\lambda$5007/H$\alpha$ = 9 
indicate [O~III] emission clouds where no H$\alpha$ was detected. }
\end{figure}

\begin{figure}
%\epsscale{0.5}
%\plotone{fig3_v2.ps}
\caption{ Barycentric velocity field of the line-emitting gas in NGC
4388.  (top) H$\alpha$; (bottom) [O~III] $\lambda$5007. Same spatial
scale and orientation as Fig. 1.  The optical continuum nucleus is
indicated in each panel by a cross. The velocites range from about
--200 km s$^{-1}$ to +200 km s$^{-1}$ relative to systemic (= 2,525 km
s$^{-1}$).}
\end{figure}

\begin{figure}
%\epsscale{0.5}
%\plottwo{fig4a.eps}{fig4b.eps}
%\plottwo{fig4c.eps}{fig4d.eps}
\caption{ Models of the disk velocity field: (top left) a simple
axisymmetric kinematic model, (top right) a kinematic model with
elliptical streaming motion in the center. See text and Table 1 for
more details on the parameters of the models. The residuals after
subtracting the axisymmetric model from the observed H$\alpha$
velocity field are presented in the bottom left panel while the bottom
right panel shows the residuals after subtracting the non-axisymmetric
model. The non-axisymmetric model is clearly a better representation
of the data.}
\end{figure}

\begin{figure}
%\epsscale{0.5}
%\plotone{fig5_v8.ps}
\caption{ Geometry of the ICM -- NGC 4388 interaction. The values of
the parameters indicated on this figure and discussed in the text are:
$i$ = -- 78$\arcdeg$, $\phi$ = 40$\arcdeg$, $\theta$ $>$ 12$\arcdeg$,
$R_{\rm H~I}$ = 10 kpc, $V_{\rm ICM}$ = 1,500 km s$^{-1}$ , $n_{\rm
ICM}$ $\sim$ 10$^{-4}$ cm$^{-3}$, $V_{\rm ps}$ = 500 km s$^{-1}$,
$n_{\rm ps}$ $\sim$ 3 $\times$ 10$^{-3}$ cm$^{-3}$. The observer is
located on the right in the same plane as the figure and at a distance
of 16.7 Mpc.}
\end{figure}

\clearpage

\setcounter{figure}{0}
\begin{figure}
%\epsscale{0.5}
\plotone{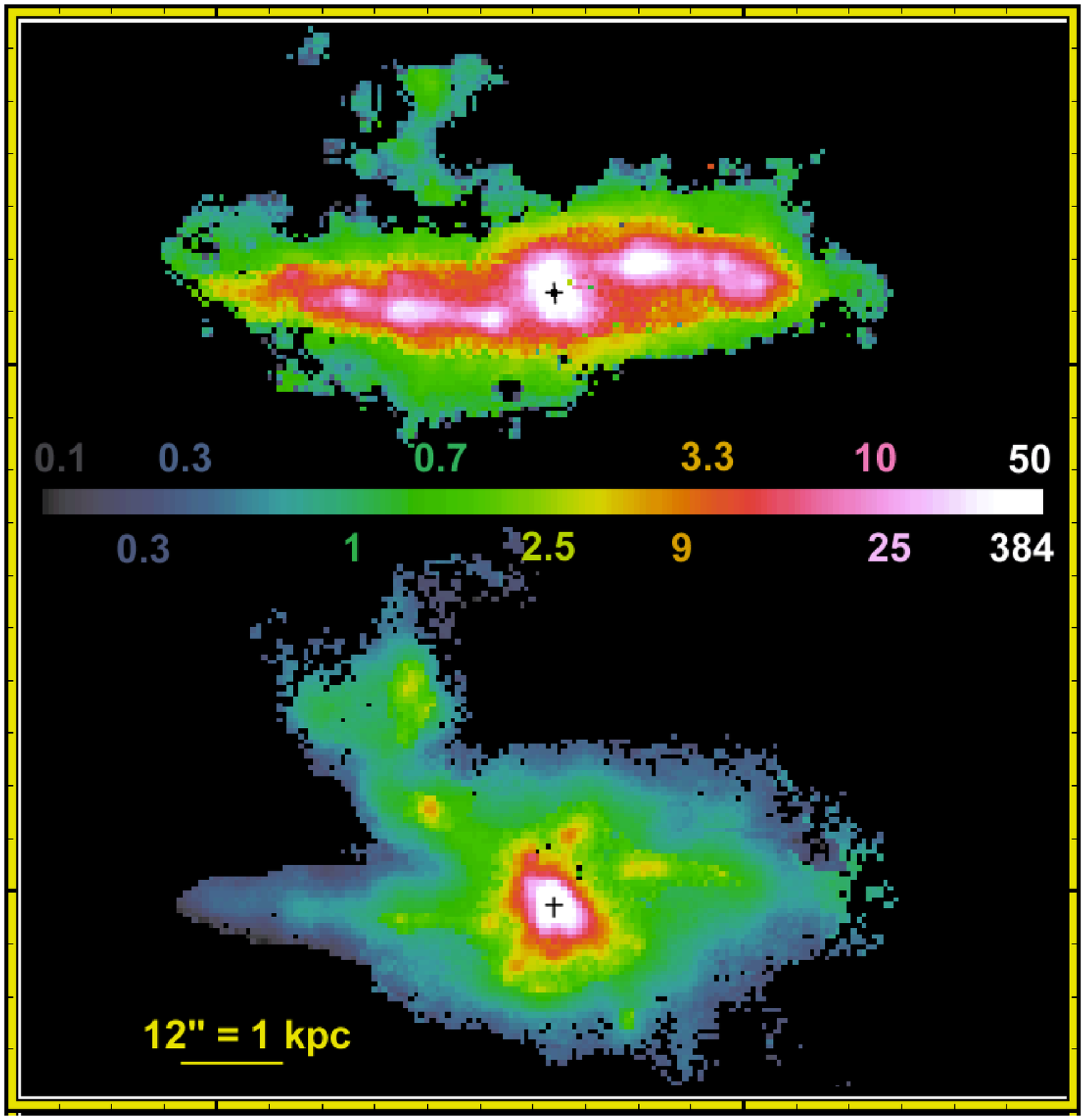}
\caption{ 
%The distributions of the line emission in NGC 4388. (top)
%H$\alpha$; (bottom) [O~III] $\lambda$5007.  The optical fluxes were
%determined from Gaussian fitting of the emission-line profiles. North
%is at the top and East to the left. The spatial scale, indicated by a
%horizontal bar at the bottom of the image, corresponds to 12$\arcsec$,
%or 1 kpc for the adopted distance of 16.7 Mpc. The optical continuum
%nucleus is indicated in each panel by a cross. The flux scale is
%logarithmic and is expressed in units of 10$^{-16}$ erg s$^{-1}$
%cm$^{-2}$ arcsec$^{-2}$.  The flux scale of the H$\alpha$ map ranges
%from 2 $\times 10^{-14}$ erg s$^{-1}$ cm$^{-2} $arcsec$^{-2}$ in the
%nucleus to $\sim$ 7$\times 10^{-17}$ erg s$^{-1}$ cm$^{-2}$ arcsec$^{-2}$,
%while the scale of the [O~III] flux map ranges from 5 $\times
%10^{-14}$ erg s$^{-1}$ cm$^{-2} $arcsec$^{-2}$ in the nucleus to $\sim$ 3
%$\times 10^{-17}$ erg s$^{-1}$ cm$^{-2}$ arcsec$^{-2}$.  
}
\end{figure}

\begin{figure}
%\epsscale{0.5}
\plotone{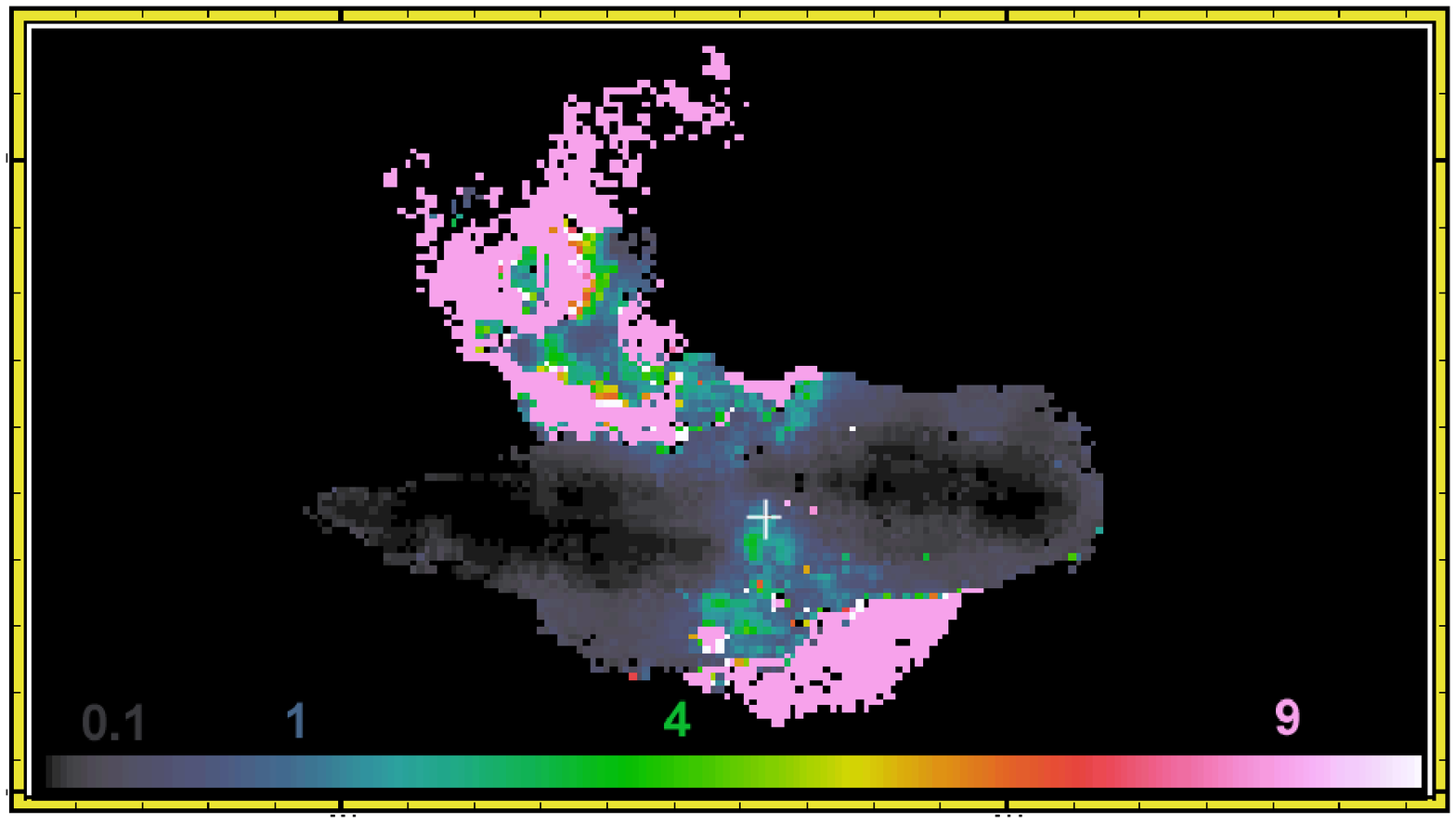}
\caption{ 
%[O~III] $\lambda$5007/H$\alpha$ flux ratio map of the
%line-emitting gas in NGC 4388. Same spatial scale and orientation as
%Fig. 1.  The optical continuum nucleus is indicated by a cross. Values
%for this ratio range from $\sim$ 0.1 in the disk to $\ga$ 5 in the
%extraplanar material. Regions with [O~III]~$\lambda$5007/H$\alpha$ = 9 
%indicate [O~III] emission clouds where no H$\alpha$ was detected. 
}
\end{figure}

\begin{figure}
%\epsscale{0.5}
\plotone{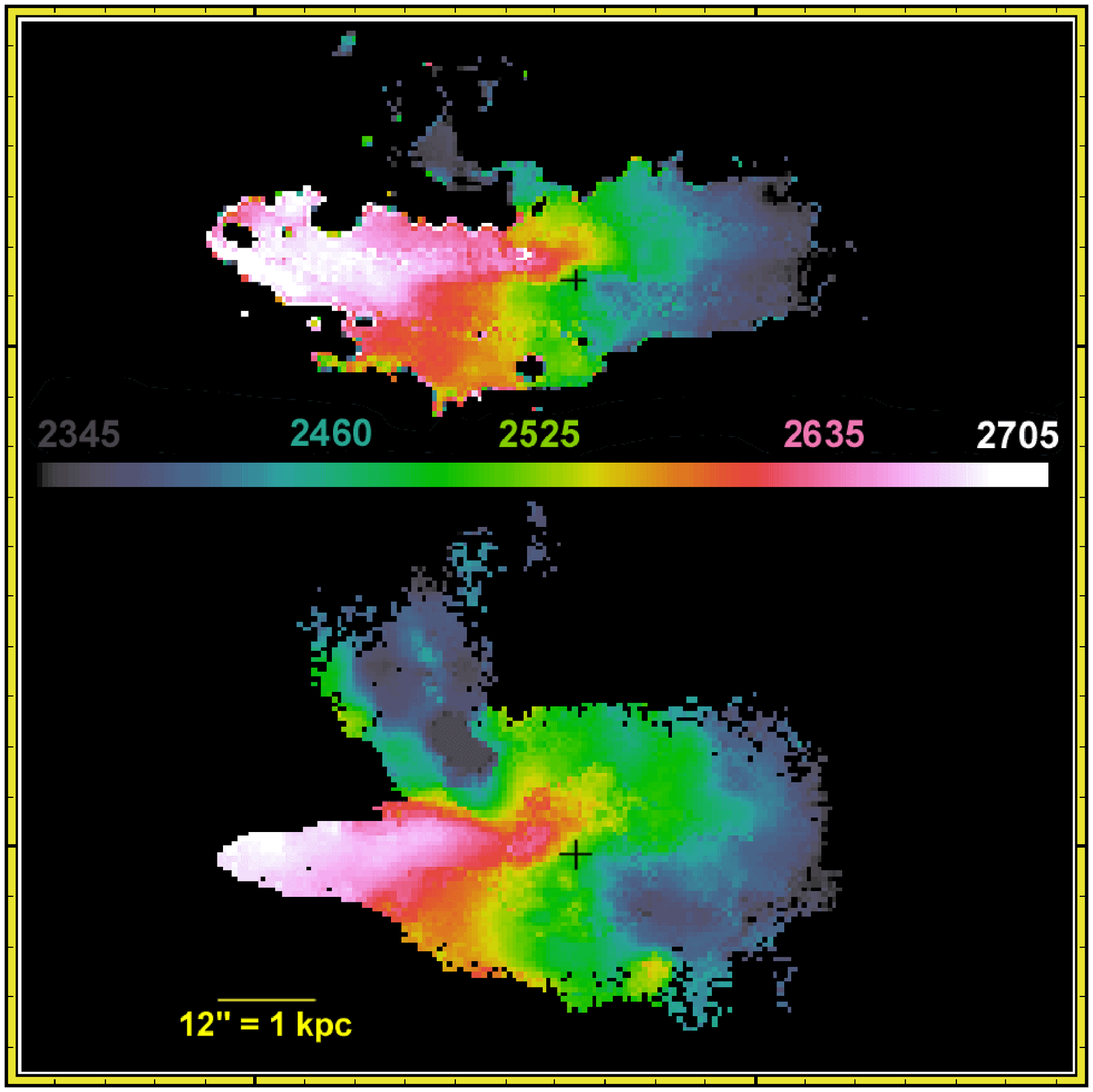}
\caption{ 
%Barycentric velocity field of the line-emitting gas in NGC
%4388.  (top) H$\alpha$; (bottom) [O~III] $\lambda$5007. Same spatial
%scale and orientation as Fig. 1.  The optical continuum nucleus is
%indicated in each panel by a cross. The velocites range from about
%--200 km s$^{-1}$ to +200 km s$^{-1}$ relative to systemic (= 2,525 km
%s$^{-1}$).
}
\end{figure}

\begin{figure}
%\epsscale{0.5}
\plottwo{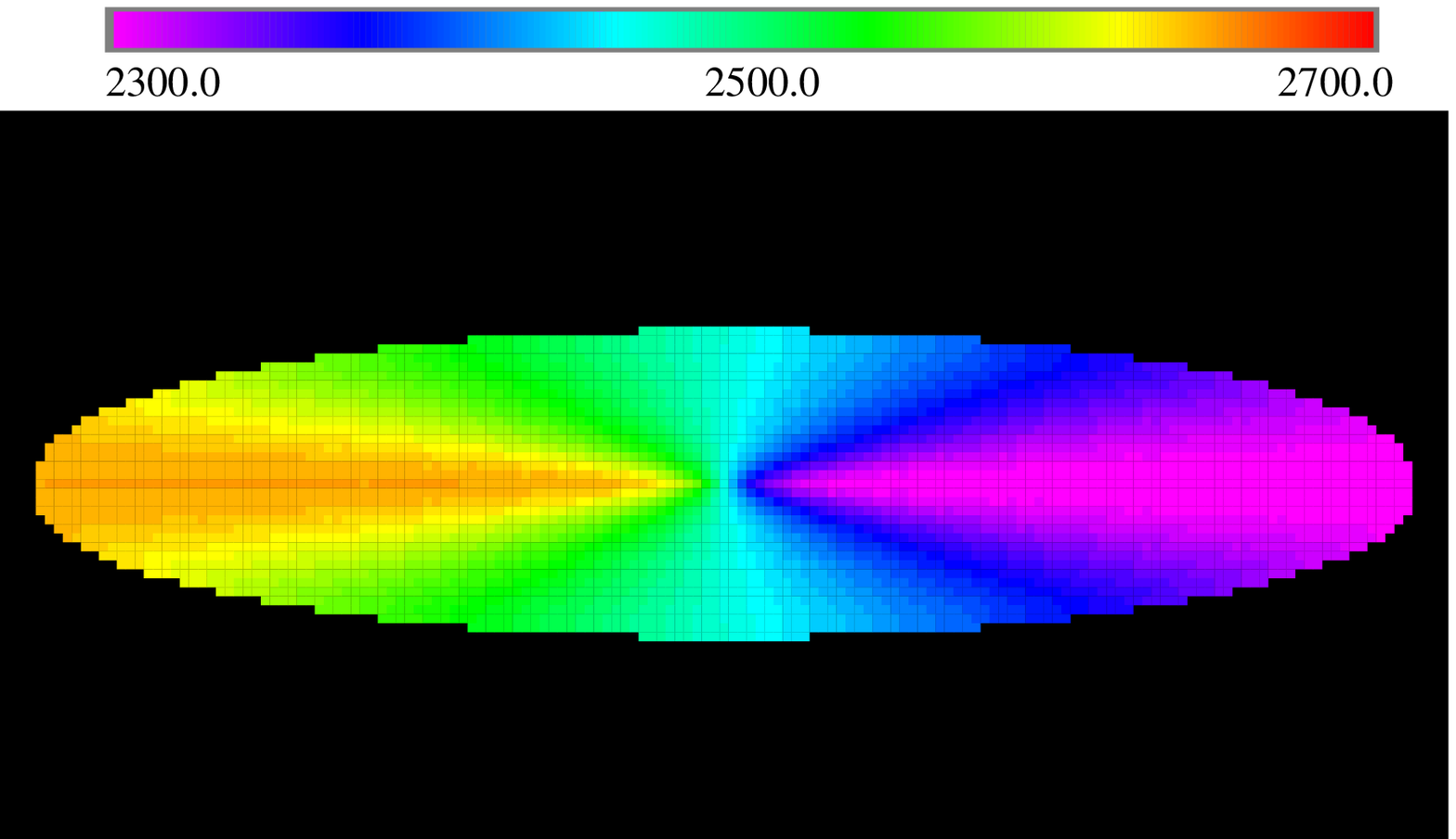}{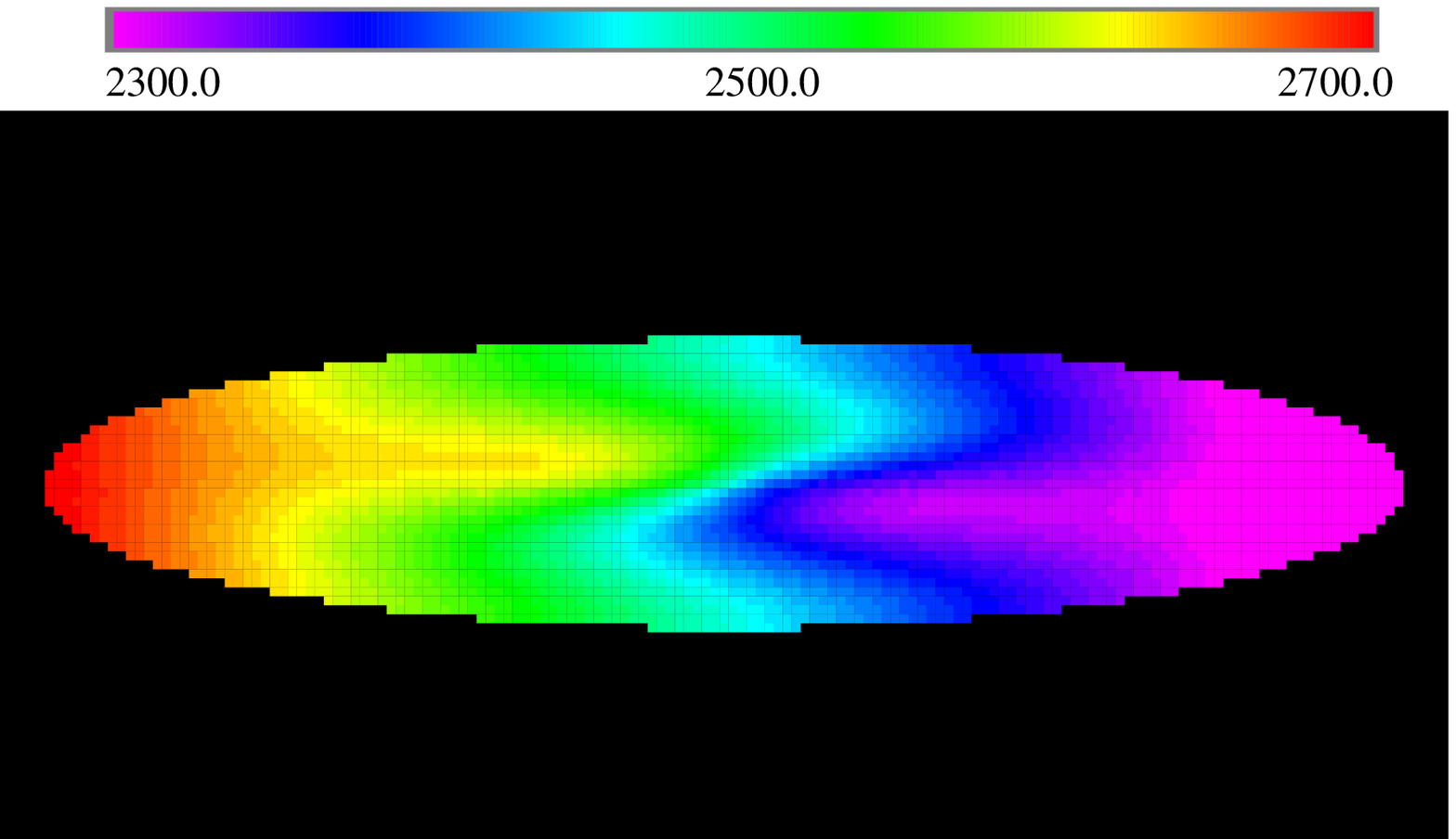}
\plottwo{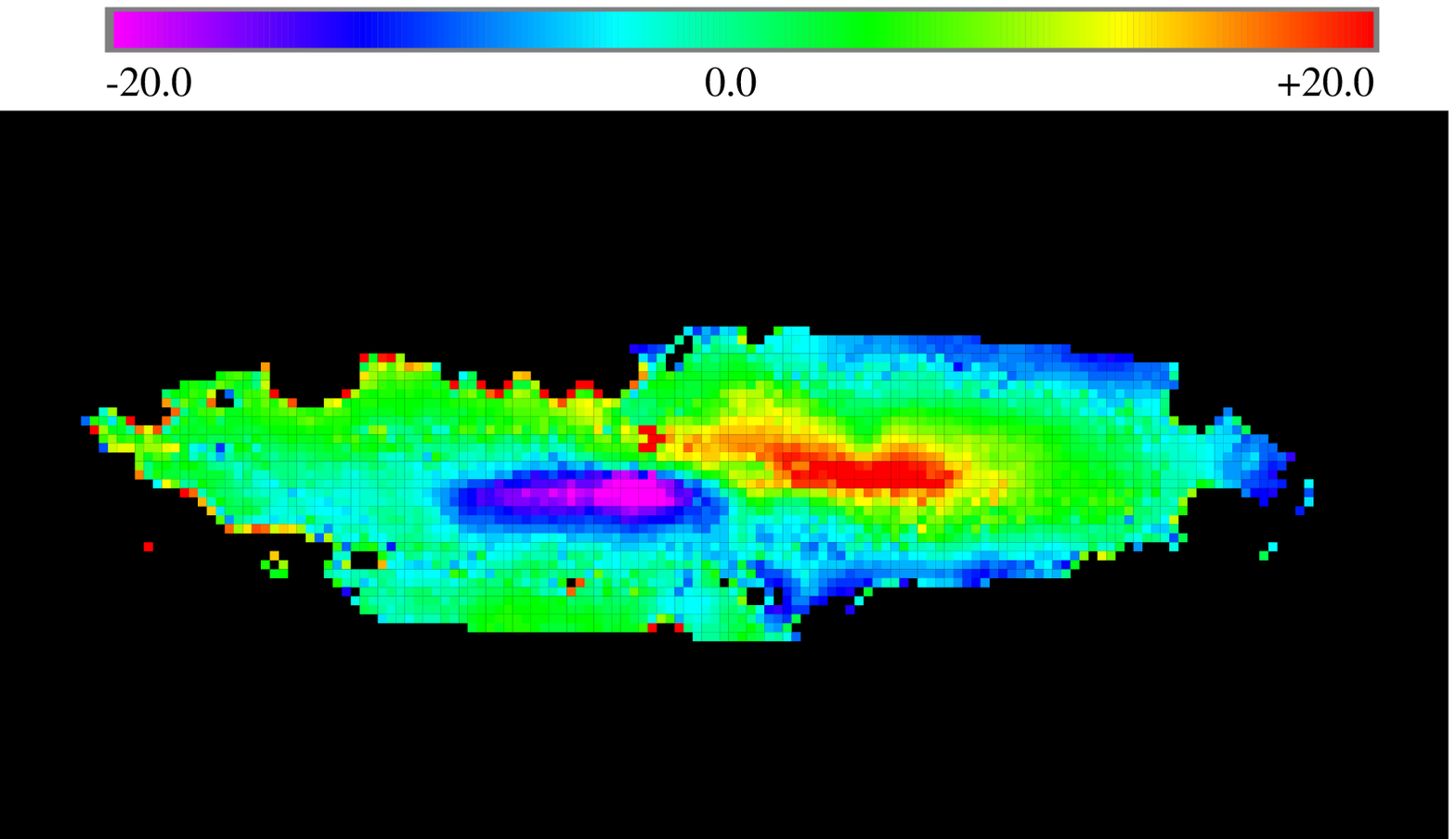}{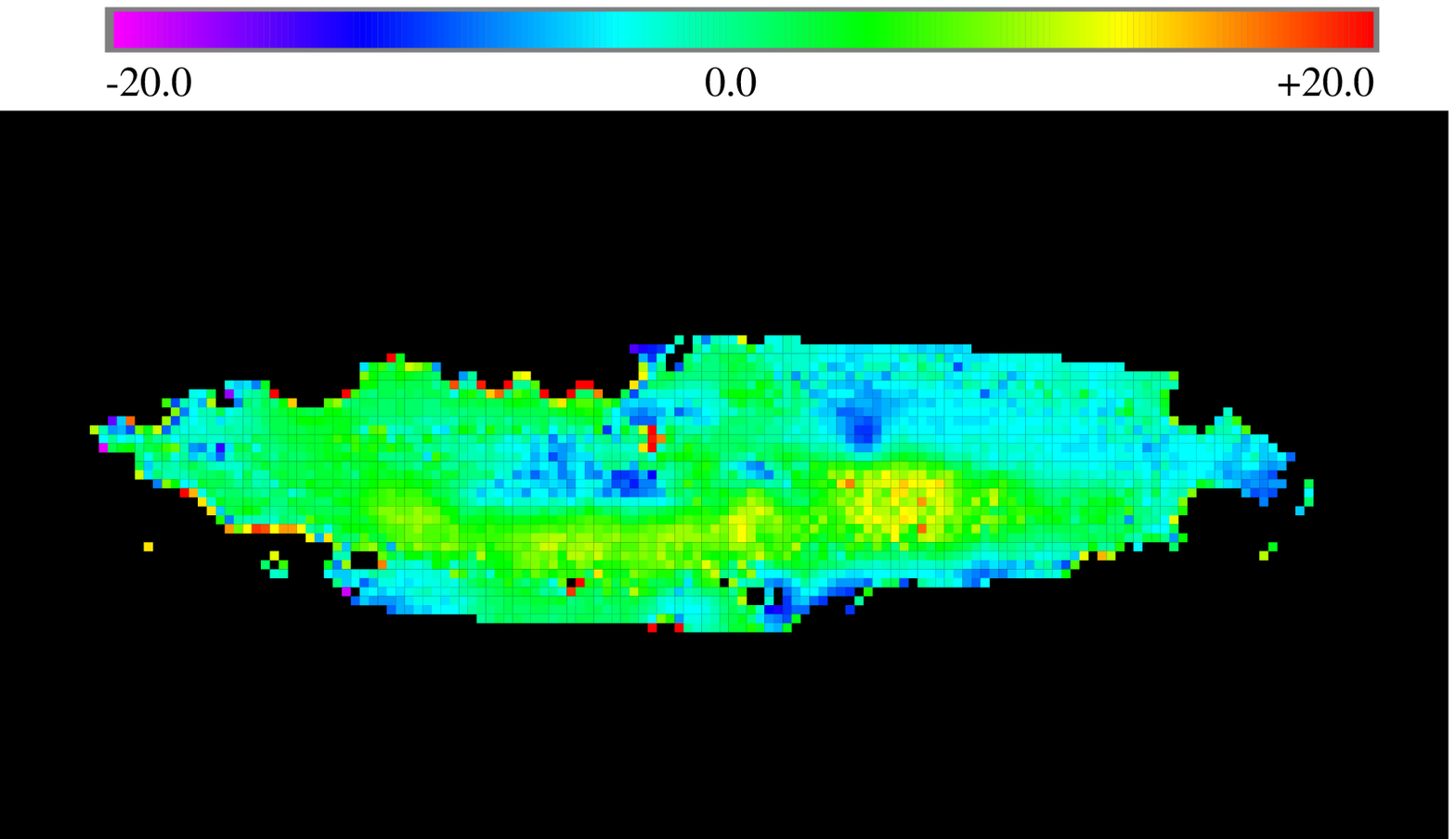}
\caption{ 
%Models of the disk velocity field: (top left) a simple
%axisymmetric kinematic model, (top right) a kinematic model with
%elliptical streaming motion in the center. See text and Table 1 for
%more details on the parameters of the models. The residuals after
%subtracting the axisymmetric model from the observed H$\alpha$
%velocity field are presented in the bottom left panel while the bottom
%right panel shows the residuals after subtracting the non-axisymmetric
%model. The non-axisymmetric model is clearly a better representation
%of the data.
}
\end{figure}

\begin{figure}
%\epsscale{0.5}
\plotone{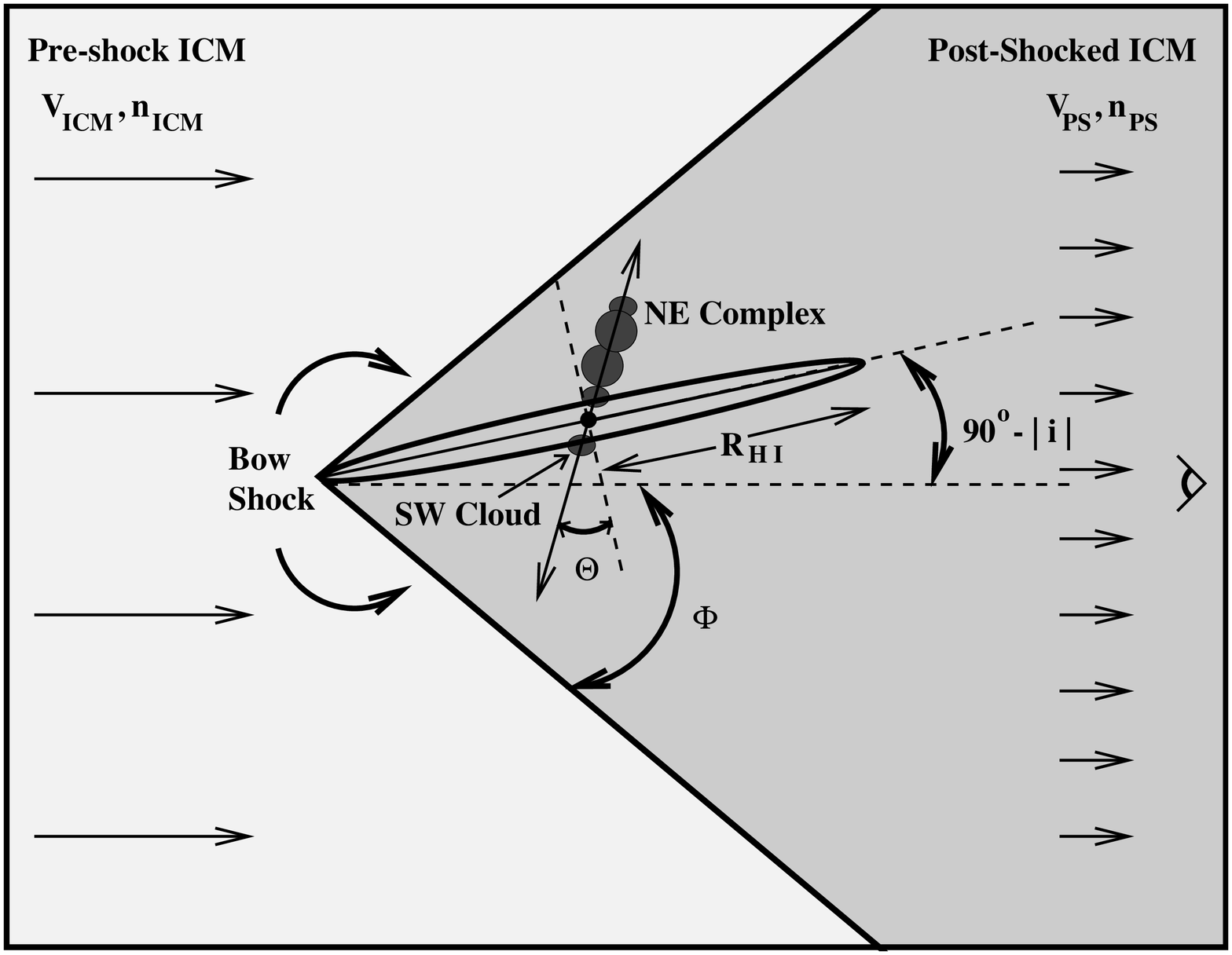}
\caption{ 
%Geometry of the ICM -- NGC 4388 interaction. The values of
%the parameters indicated on this figure and discussed in the text are:
%$i$ = -- 78$\arcdeg$, $\phi$ = 40$\arcdeg$, $\theta$ $>$ 12$\arcdeg$,
%$R_{\rm H~I}$ = 10 kpc, $V_{\rm ICM}$ = 1,500 km s$^{-1}$ , $n_{\rm
%ICM}$ $\sim$ 10$^{-4}$ cm$^{-3}$, $V_{\rm ps}$ = 500 km s$^{-1}$,
%$n_{\rm ps}$ $\sim$ 3 $\times$ 10$^{-3}$ cm$^{-3}$. The observer is
%located on the right in the same plane as the figure and at a distance
%of 16.7 Mpc.
}
\end{figure}

\end{document}